\documentclass[reprint, aps,prb,eqsecnum,showpacs,twocolumn,floats,superscriptaddress,10pt]{revtex4-1}
\usepackage{graphicx}
\usepackage{dcolumn}
\usepackage{amsmath}
\usepackage{verbatim}
\usepackage{subeqnarray}
\usepackage{bm}
\usepackage{color}
\usepackage[caption=false]{subfig}
\usepackage{amssymb}
\usepackage{mathrsfs}
\usepackage[breaklinks=true,colorlinks=true,linkcolor=blue,urlcolor=blue,citecolor=magenta]{hyperref}

\begin{document}
\title{Periodically Driven Three-Level Systems}
\author{M. B. Kenmoe}
\affiliation{Mesoscopic and Multilayer Structures Laboratory, Faculty of Science, Department of Physics, University of Dschang, Cameroon}
\author{L. C. Fai}
\affiliation{Mesoscopic and Multilayer Structures Laboratory, Faculty of Science, Department of Physics, University of Dschang, Cameroon}
\date{\today}
\begin{abstract}
We study the dynamics of a three-level system (ThLS) sinusoidally driven in both longitudinal and transverse directions  and in the presence of a uniaxial anisotropy $D$ entering the generic Hamiltonian through the zero-energy splitting term $D(S^{z})^{2}$ where $S^{z}$ is the projection of the spin vector along the quantization direction. As a consequence of the addition of this term,  the order of the symmetry group of the Hamiltonian is increased by a unit and we observe a sequence of cascaded $SU(3)$ Landau-Zener-St\"uckelberg-Majorana (LZSM) interferometers.  The study is carried out  by analytically  and numerically calculating the probabilities of non-adiabatic and adiabatic evolutions. For non-adiabatic evolutions, two main approximations based on  the weak and strong driving limits are discussed by comparing the characteristic frequency of the longitudinal drive with the amplitudes of driven fields.  For each of the cases discussed, our analytical results quite well reproduce the gross temporal profile of the exact numerical probabilities. This allows us to check the range of validity of analytical results and confirm our assumptions.  For adiabatic evolutions, a general theory is constructed allowing for the description of adiabatic passages in arbitrary ThLSs in which direct transitions between states with  extremal spin projections are forbidden. A compact formula for adiabatic evolutions is derived and numerically tested for some illustrative cases. Interference patterns demonstrating multiple LZSM transitions are reported.  Applications of our results to the Nitrogen Vacancy Center (NVC) in diamond are discussed. 
\end{abstract}
\maketitle

\section{Introduction}
One of the prerequisites for producing quantum interferences is the ability to generate states  energy that compulsorily cross or come close at least twice in the course of variation of a control parameter such that a quantum system following the corresponding paths splits into two separated waves at first crossing (splitter) and recombining into a single wave in the next crossing (mixer)\cite{Gefen, Hanggi, Kayanuma1994, Garraway}. This in general induces accumulation of dynamical phases and superposition of wave functions leading thus to  the formation of fringes that are inspected and probed in spectroscopic  analyzes to capture some information  relevant to the complex dynamics of a system bathing in its environment\cite{Schaw, Thompson, Oliver}. However, two crossings (double passage) might not be enough to produce desired information (an exceptional case of quantum interferences with a double-crossing configuration is discussed in Ref.\onlinecite{Mark}), instead, some of the control parameters of the system are more often periodically changed such that the system oscillates back and forth around an avoided level crossing\cite{Sun, Jian}. 

This procedure has already been applied to two-level systems (TLSs) with great predictions\cite{Gefen, Hanggi, Kayanuma1994, Garraway}. For these reasons and due to versatile  applications in quantum technology, cryptography and metrology\cite{Ekert, Knill, Duan, Scarani}, huge and remarkable alike attentions, emerging both from theoretical and experimental viewpoints are currently granted to qubit, the unit of Quantum Information Processing (QIP)\cite{Josephson, QDots, Devoret, Qbreview, Brien}. The former (qubit) is unquestionably ubiquitous in nature and gently imposes as a promising candidate for the breakthrough in developing quantum technologies. For the successful realization of the future quantum computer, superconducting qubits comprising charge and flux qubits are implemented in various solid-state real or artificial  devices, ranging from Josephson junctions\cite{Josephson} to quantum dot based devices\cite{QDots}. Features  of qubit are inferred for instance by reading out its response to external perturbations or by periodically varying some of its control parameters (see Ref.\onlinecite{Shev} for review). The last technique led to the uncovering of interference patterns associated with Landau-Zener-St\"uckelberg-Majorana (LZSM) oscillations\cite{lan, zen, stu, Maj} in population of levels allowing to decipher and explore complex dissipative effects on the qubit and to measure some relevant parameters such as coherence time and inhomogeneous decay time\cite{Shytov, Ludwig}. An experimental method based on steady-state allowing to collect all relevant information about a qubit is found in Ref.\onlinecite{Ludwig}. 

In QIP, the storage of information is based on quantum logic gates that are mostly made of qubits, but can also be composed of qutrits\cite{Liu, Lan}. The qutrit is the unit of information in ternary quantum computing made of  the superposition of three states\cite{Liu, Lan}. It is the quantum analogous of the classical trit.
Within the vast and incommensurable literature devoted to the QIP, few analytic studies address the questions of qutrits coherently controlled by an electromagnetic field or periodically driven\cite{Zhou, Du,  comment1, Ansari}. This is most likely due to the complexity of handling  ThLS' dynamics theoretically and/or the difficulty of reading them out experimentally. Indeed, current attempts/proposals in this regard tend to simplify the complex dynamics of the ThLS to that of a TLS which is easier to handle\cite{Zhou, Du,  comment1, Ansari, Fuchs, Wubs, Danon, Barfuss}. For instance, by applying an additional static magnetic field,  the degeneracy is lifted between states. 

 As compared to qubit, the qutrit is more robust against environmental effects, encodes more information, has a longer coherence time, can already be implemented at room temperature\cite{Lan, Dolde} and provides better security in QIP\cite{Lan, Langford}. For these reasons, the qutrit plays a crucial role in QIP. To process information carried, qutrits should preserve their coherent states for a longer time than is necessary to lose the information encoded,  and ideally they must be completely isolated from unwanted external influences. Qutrits  unfortunately are found in host exotic and quantum devices (three-well potentials\cite{Zhou},  quantum electrodynamics circuits\cite{Liu},  optical lattice\cite{Medendorp}, Nitrogen Vacancy Centers\cite{Zu} (NVC), superconducting quantum circuits\cite{Bian, Zhou}  etc) where nuclear-spin interactions and molecular dynamics prevail and may have drastic incidence on their coherence. Due to the difficulty of isolating a qutrit from its environment, along with the inability to realize practical quantum systems that operate with a large number of qutrits, currently realizable quantum computers can only be of small number of qubits/qutrits\cite{Seth, Kane, Monz}.  These observations raise the key question of qutrit coherent dynamic control which can be achieved  by switching the detuning between finite values and zero at speeds ranging from the regime of non-adiabatic (fast sweep) to that of adiabatic evolutions (slow sweep) or by applying a longitudinal and/or a transverse electromagnetic field.

In this paper, we study the dynamics of a ThLS periodically driven in both longitudinal and transverse directions and in the presence of an easy-axis anisotropy $D$.  We investigate the possibility of performing a coherent control of ThLSs by simultaneously applying two classical fields with respect to a quantization axis.  Nuclear-spin exchanges causing hyperfine interactions as in quantum dots based devices are, however, neglected.   Due to the difficulty of analytically solving the relevant Schr\"odinger equation in an exact basis\cite{comment}, (the latter is exactly solved numerically) two main approximations are made: the transverse and longitudinal drive approximations. In both cases, we consider the two complementary limits of weak and strong drives obtained by comparing the amplitude (in frequency units since $\hbar=1$) of the driving fields with the characteristic frequency of the longitudinal field. We open a deeper perspective of using spectroscopic analysis to probe ThLSs' dynamics by periodic drives and the possibility of discussing the intrinsic dynamics of three-level spin-$1$ systems without necessarily   reducing their dynamics to that of an approximate TLS. We believe that this opens a new route to realizing optimal control of ThLSs for future implementation in quantum computers.

The paper is organized as follows: in Sec.\ref{model}, the model of the study is presented and its applications are discussed. In Secs.\ref{Sec1} and \ref{Sec2}, the transverse and longitudinal driving approximations are discussed. In Sec.\ref{Sec3}, more approximations also useful for experimental applications are implemented. A possible application of our results to  QIP is discussed in Sec.\ref{Sec4}.  The paper concludes in Sec.\ref{Sec5} with a summary of our main achievements.

\section{Model}\label{model}
We consider a ThLS coherently driven in both longitudinal and transverse directions with time-dependent periodic classical fields (one can see Ref.\onlinecite{Jian} for a similar discussion on qubits). In the longitudinal direction, the field is of amplitude $A$ and frequency $\omega$ (a zero-phase field at initial time). It is also detuned in that direction by an easy-axis anisotropy $D$ (static part of the detuning). In the transverse direction, one applies another periodic field now of amplitude $\mathcal{A}_{f}$ and frequency $\omega _{f}$.  Throughout the paper, the reduced Planck's constant $\hbar=1$; therefore, the amplitudes of the fields are of the dimension of frequency and comparable with $\omega$ and/or $\omega_{f}$. This fact will be shortly exploited. The  setup is described by the prototype Hamiltonian 
\begin{equation} \label{equ1} 
\mathcal{H}(t)=\mathcal{H}_{Q}(t)+\mathcal{H}_{drive}(t)+D(S^{z})^{2}, 
\end{equation}
where the first term $\mathcal{H}_{Q}(t)=A\cos (\omega t)S^{z}$ is the Hamiltonian of the longitudinal frequency-modulated ThLS and the second term $\mathcal{H}_{drive}(t)=\mathcal{A}_{f}\cos(\omega _{f} t)S^{x}$ stands for the longitudinal field-induced  Hamiltonian  which pours into the system, the energy necessary to excite its states. The last term describes a zero-field energy splitting between the  middle state and its neighbors.  $S^{\nu}$ with $\nu=x,y,z$ are spin operators or, more  specifically, the generators of the  $SU(2)$ group associated with the Lie algebra $su(2)$ here given by the commutation relations $[S^{\mu},S^{\nu}]=i\epsilon_{\gamma}^{\mu\nu}S^{\gamma}$ where $\epsilon_{\gamma}^{\mu\nu}$ are structure constants on $SU(2)$ and $[A,B]=AB-BA$ (see Ref.\onlinecite{Gell}).  Though $\mathcal{H}(t)$ is expressed in terms of generators of the Lie group $SU(2)$, it embeds the hidden dynamical symmetry of the Lie group $SU(3)$. We observe that the $SU(2)$ symmetry is broken down by adding the term $D(S^{z})^{2}$ which renders $\mathcal{H}(t)$ nonlinear in $SU(2)$. As shown in Ref.\onlinecite{Ken2013}, this seemingly non-linearity is removed in the Lie group $SU(3)$ where in turn, $\mathcal{H}(t)$ becomes a trajectory and expresses as a linear combination of Gell-mann matrices\cite{Ken2013} (generators of the $su(3)$ algebra\cite{Gell}). 

\begin{figure}[]
\begin{center}
\includegraphics[width=9cm, height=4cm]{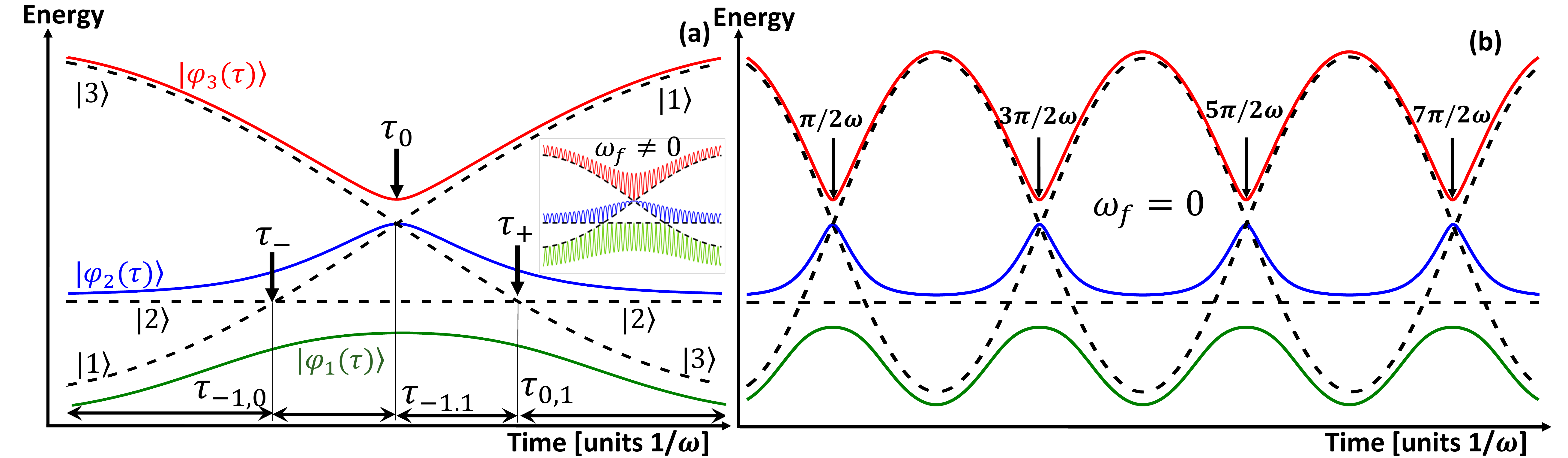}
\end{center}
\vspace{-0.5cm}
\caption{ (Color Online) Sketch of energy diagrams associated with the model Eq.(\ref{equ1}). The left panel is a full cycle of the $SU(3)$ LZSM interferometry and the right panel indicates periodic trajectories. Dashed lines are energies of diabatic states while solid lines are adiabatic energies of the Hamiltonian when the frequency of the transverse drive is set to zero ($\omega_{f}=0$). Inserted is a full cycle of the $SU(3)$ LZSM interferometry for non-zero and large $\omega_{f}$. It  appears from insert that adiabatic trajectories oscillate, and this, rapidly. As $\omega_{f}$ increases, adiabatic states can interfere/touch at avoided crossings such that information can be directly transferred from one adiabatic state to another. This is a clear indication that in the regime of strong transverse drives, corresponding to adiabatic-like drive, one may expect quantum interferences as $\omega_{f}$ increases.  In addition to the $SU(2)$ LZSM model, this is yet another situation in which adiabatic principle of quantum mechanics breaks down resulting in transitions between adiabatic states. The time $\tau$ is in the unit of $1/\omega$.}
\label{Figure0}
\end{figure}

As a starting point of our description, let us introduce an appropriate basis for spin operators. We assume for a while that the transverse drive is completely switched off ($\mathcal{A}_{f}=0$). Thereof, the eigenstates of $\mathcal{H}(t)$ are eigenstates of $\mathcal{H}_{Q}(t)$. They match the $2+1$ projections $m_{S}=\pm1,0$ of the total spin vector along the quantization direction in the absence of anisotropy ($D=0$). They are called diabatic states and denoted as $|m_{S}=-1\rangle$, $|m_{S}=0\rangle$ and $|m_{S}=1\rangle$ respectively for the lower, middle and upper levels. Spin operators are written in that basis as $S^{z}=|m_{S}=1\rangle\langle m_{S}=1|-|m_{S}=-1\rangle\langle m_{S}=-1|$ and $S^{x}=(|m_{S}=0\rangle\langle m_{S}=-1|+|m_{S}=-1\rangle\langle m_{S}=0|+|m_{S}=0\rangle\langle m_{S}=1|+|m_{S}=1\rangle\langle m_{S}=0|)/\sqrt{2}$ and the  Hamiltonian acquires the form\cite{comment3}
\begin{eqnarray} \label{equ2} 
\mathcal{H}(t)=
\left[
{\begin{array}{*{20}c}
A\cos (\omega t)+D & \frac{\mathcal{A}_{f}}{\sqrt{2}}\cos (\omega _{f} t) & 0\\
\frac{\mathcal{A}_{f}}{\sqrt{2}}\cos (\omega _{f} t) & 0 & \frac{\mathcal{A}_{f}}{\sqrt{2}}\cos (\omega _{f} t)\\
0  & \frac{\mathcal{A}_{f}}{\sqrt{2}}\cos (\omega _{f} t) & -A\cos (\omega t)+D
\end{array} } \right].\quad
\end{eqnarray}
Diabatic eigen-energies $\mathcal{E}_{m_{S}}(t)=Am_{S}\cos (\omega t)+Dm^{2}_{S}$ are plotted in the figure \ref{Figure0} (dashed lines). The system undergoes sudden transitions at times $t_{m_{S},m_{S}'}=\arccos(-[m_{S}+m_{S}']D/A)/\omega+2\pi N/\omega$ (where $N=0,1,2,3,...$) where two diabatic energies $\mathcal{E}_{m_{S}}(t)$ and $\mathcal{E}_{m_{S}'}(t)$ cross. The corresponding curves  indicate non-adiabatic trajectories followed by the ThLS during the fast drive.

For $N=0$, the Hamiltonian $\mathcal{H}(t)$ describes a full cycle of the $SU(3)$ LZSM interferometry\cite{Ken2013} (see the corresponding energy diagram displayed in figure \ref{Figure0}(a)). Three crossing regions can be identified: one at time $t_{-1,0}\equiv \tau_{-}$, another at $t_{-1,1}\equiv \tau_{0}$, and the last one at $t_{0,1}\equiv \tau_{+}$. This shows that if the ThLS is prepared for instance in the state $|2\rangle$ at an initial time far from the left of $\tau_{-}$, it splits twice: once at $\tau_{-}$ (first splitter) and another at $\tau_{0}$ (second splitter), and recombines at $\tau_{+}$ (mixer) before arriving any of the other states at time $t$ far from the right of $\tau_{+}$. These interactions between quantum paths lead to LZSM interference patterns\cite{Ken2013} (a prototype is presented in Fig.\ref{Figure2bis}) that can be destructive or constructive (see Ref.\onlinecite{Shev} for ample discussion). Such a splitting and recombination of populations can also be observed with other preparations of the ThLS. 

 For $N\neq0$, the model describes a sequence of $SU(3)$ LZSM interferometers.   Therefore, by periodically changing one of the control parameters (detuning and/or Rabi coupling) of a single $SU(3)$ LZSM interferometry, the resulting system behaves as a combination of multiple such interferometers (see Fig.\ref{Figure0}(b)). The double $SU(3)$ LZSM interferometry realizable by restricting periodic passages through two cycles or by only coupling two interferometers, is predicted in Ref.\onlinecite{Gallego} using a triple quantum dot chain configuration (coupled LZSM quantum dot interferometers). 

  Considering interplay between quantum paths, we note that direct transitions between the states with extremal spin projections are forbidden. The resonant transverse field in $\mathcal{H}_{drive}(t)$ only ensures the pair of transitions $|m_{S}=-1\rangle\leftrightarrow |m_{S}=0\rangle$ and $|m_{S}=0\rangle\leftrightarrow |m_{S}=1\rangle$ while direct transitions  $|m_{S}=-1\rangle\leftrightarrow |m_{S}=1\rangle$ are not possible. Transitions are mediated by tunneling between neighboring states. The system passes through the intermediate state $|m_{S}=0\rangle$.

Furthering the description of the features of our model, we must realize that when the $SU(3)$ symmetry breaks down by suppressing the easy-axis anisotropy ($D=0$), all diabatic states cross at a single point in one cycle. There is a unitary operation  $\mathbf{R}=e^{-i\pi S^{y}/2}$ which transforms $S^{z}\to S^{x}$  and $S^{x}\to-S^{z}$ subsequently inter-changing the actions of the fields such that $A\to\mathcal{A}_{f}$, $\mathcal{A}_{f}\to-A$ and $\omega\rightleftharpoons\omega_{f}$. Then for instance, if the dynamics of the system driven by the transverse field is known when the longitudinal field is turned off, the inverse situation (dynamics with longitudinal drive in the absence of the transverse drive) is easily probed as deductible from the formal case by reverse engineering with $\mathbf{R}$.	This is a consequence of the choice we have made to ascribe the same shape of the signal (cosine shape) to the longitudinal and transverse drives. 

From now onwards, we consider and discuss four limiting cases adopted by comparing the characteristic frequency $\omega$ of the longitudinal drive with $A$ and $\mathcal{A}_{f}$ that are of frequency dimension as $\hbar=1$. Firstly, we compare $\omega$ and $\mathcal{A}_{f}$ (transverse drive approximations) and distinguish between the limits of weak ($\mathcal{A}_{f}\ll\omega$) and strong ($\mathcal{A}_{f}\gg\omega$) transverse drives. In the second case $\mathcal{A}_{f}\gg\omega$, diabatic states (eigenstates of $\mathcal{H}_{Q}(t)$) hybridize at level crossings rather forming avoided level crossings (solid lines on figure \ref{Figure0}). The eigenstates of the Hamiltonian in this situation are called adiabatic states and define the trajectories followed by the system during slow changes-in-time of the Hamiltonian. Secondly, we proceed similarly by comparing $\omega$ and $A$ (longitudinal drive approximations). We equally distinguish between the limits of weak ($A\ll\omega$) and strong ($A\gg\omega$) longitudinal drives. Such limiting cases may be relevant for future applications in Bose Einstein Condensate\cite{Mark}, quantum computing\cite{Ekert}, symmetric double-well potential\cite{Schatzer}, NVC\cite{Huang}, atoms in optical lattices\cite{Kling} etc.

\section{Transverse driving approximations}\label{Sec1}

We are typically interested in the time-evolution of populations in diabatic levels during variations of external fields (longitudinal and transverse drives). However, such populations strongly depend on how the fields are tuned. We mainly focus on two complementary regimes of the  transverse field: the weak transverse limit, treated in the diabatic basis as described in detail below, and the extreme limit of the strong transverse  drive elucidated in the adiabatic basis.

\subsection{Weak transverse driving limit $\mathcal{A}_{f}\ll\omega$}\label{diab}

This limit is addressed in the basis of the eigenstates of $\mathcal{H}_{Q}(t)$.  Let $|\Psi (t)\rangle$ be the total wave-function of the ThLS spanned in the three-dimensional Hilbert space $\mathscr{H}$ by the basis vectors $|m_{S}=-1\rangle\equiv|1\rangle=[1,0,0]^T$, $|m_{S}=0\rangle\equiv|2\rangle=[0,1,0]^T$ and  $|m_{S}=1\rangle\equiv|3\rangle=[0,0,1]^T$ (where $T$ indicates hereafter the vector transposed) that are mutually orthogonal $\langle \kappa'|\kappa\rangle=\delta_{\kappa\kappa'}$ and satisfy the  closure relation $\sum_{\kappa=1}^{3}|\kappa\rangle\langle\kappa|=1$, ($\kappa=1,2,3$). In the Hilbert space $\mathscr{H}$, the vector $|\Psi (t)\rangle$ lies on the surface $\mathcal{S}^{2}$ of the three-dimensional sphere $\mathcal{S}^{3}$ and expresses as the linear combination 
\begin{equation} \label{equ3} 
|\Psi (t)\rangle=\sum_{\kappa=1}^{3}C_{\kappa}(t)|\kappa\rangle,
\end{equation}
where $C_{\kappa}(t)=\langle \kappa|\Psi(t)\rangle\in\mathbb{C}$. The coefficients of the expansion, are projections of $|\Psi(t)\rangle$ onto the direction of the basis vector $|\kappa\rangle$ and are subjected to the constraint $\sum_{\kappa=1}^{3}|C_{\kappa}(t)|^{2}=1$ (conservation law). From a quantum mechanical view point, $C_{\kappa}(t)$ are probability amplitudes for measuring/observing the ThLS in the state $|\kappa\rangle$. Quantum mechanics then tells us that populations we are looking for are $|\langle \kappa|\Psi(t)\rangle|^{2}$  when $|\Psi(t)\rangle$ obeys the first order linear differential equation
\begin{equation}\label{equ4} 
i\frac{d}{dt}|\Psi (t)\rangle=\mathcal{H}(t)|\Psi (t)\rangle,
\end{equation}
known as the time-dependent Schr\"odinger equation (TDSE). This equation is subjected to the initial condition $|\Psi (t_{0})\rangle=|\kappa'\rangle$. The wave-function $|\Psi (t)\rangle$ shares the same symmetry operations as $\mathcal{H}(t)$. Indeed, topologically,  the group $G_{su(3)}=e^{i su(3)}$ generated from the algebra by exponentiation is a compact manifold\cite{Victor}. The Lie algebra $su(3)$ can be considered as a three-dimensional vector space  spanned by the Gell-mann matrices in which the  Hamiltonian is a trajectory. By conjecture, the wave-function arises via exponentiation of the Hamiltonian. Then, because of the stated  topology, the wave-function is a trajectory in the Lie group $SU(3)$ while the Hamiltonian is a trajectory in the Lie algebra $su(3)$. These observations reveal that there are some locally hidden dynamical symmetries between probability amplitudes that follow from symmetries of the Hamiltonian. This idea is widely shared in this work and such hidden symmetries are elucidated and exploited to reduce the length of our calculations. 

\begin{figure}[]
\vspace{-0.25cm}
\begin{center}
\includegraphics[width=7.5cm, height=6.5cm]{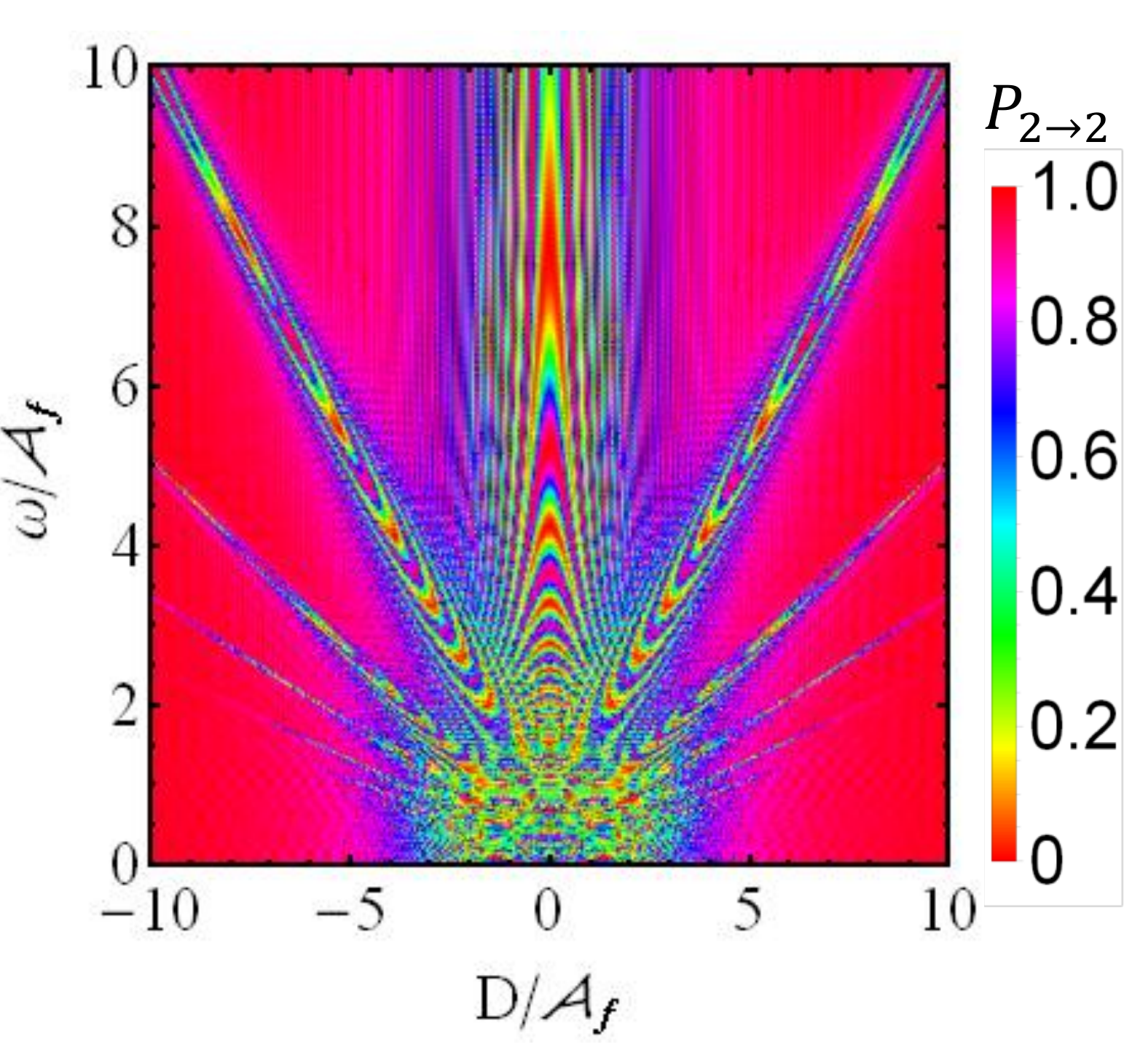} 
\end{center}
\vspace{-0.75cm}
\caption{(Color Online) Population  $P_{2\to2}(t)=|C_{2}(t)|^{2}$  on the diabatic state $|2\rangle$ at time $t=50/\mathcal{A}_{f}$ for an initial preparation of the ThLS in the state $|2\rangle$ at $t_{0}=0.0$. It is calculated by numerically solving the TDSE (\ref{equ5}) with the model (\ref{equ1}). For numerical implementation,  we have considered $A/\mathcal{A}_{f}=2.5$ and $\omega_{f}/\mathcal{A}_{f}=0.0$. We have observed that under the same condition but with rather large $\omega_{f}/\mathcal{A}_{f}$, such interference patterns appear twice (see Fig.\ref{Fig4}). }\label{Figure2bis}
\end{figure}

In the basis $\{|1\rangle$, $|2\rangle$, $|3\rangle\}$, equation (\ref{equ4}) yields a set of three coupled equations for probability amplitudes. Constructing the three-component vector $\mathbf{C}(t)=[C_{1}(t),C_{2}(t),C_{3}(t)]^{T}$, the TDSE (\ref{equ4}) becomes 
\begin{eqnarray} \label{equ5} 
i\frac{d}{dt}C_{n}(t) =\sum_{\kappa=1}^{3}\mathcal{H}_{n\kappa}(t)C_{\kappa}(t),
\end{eqnarray} 
 where $\mathcal{H}_{n\kappa}(t)$ are matrix elements of $\mathcal{H}(t)$ in (\ref{equ2}). Equation (\ref{equ5}) is solved with the initial condition $C_{\kappa'}(t_{0})=\delta_{\kappa'\kappa}$ assuming the system initialized at time $t_{0}$ in the  diabatic state $|\kappa'\rangle$. Afterward, we compute the probability $P_{\kappa'\to\kappa'}(t)\equiv P_{\kappa'}(t)=|C_{\kappa'}(t)|^{2}$ to stick in the same diabatic state $|\kappa'\rangle$ after passing the degeneracy points and the probabilities $P_{\kappa'\to\kappa}(t)\equiv P_{\kappa}(t)=|C_{\kappa}(t)|^{2}$ to be measured in a different diabatic state $|\kappa\neq\kappa'\rangle$ after interactions. For now, this task is numerically performed for an initial occupation of $|2\rangle$ (see figure \ref{Figure2bis}). Associated  interference patterns corresponding to interactions between intermediate paths from $|2\rangle$ to $|2\rangle$ is reported. 

\begin{figure}[]
\vspace{-0.5cm}
\begin{center}
\includegraphics[width=4.2cm, height=3.5cm]{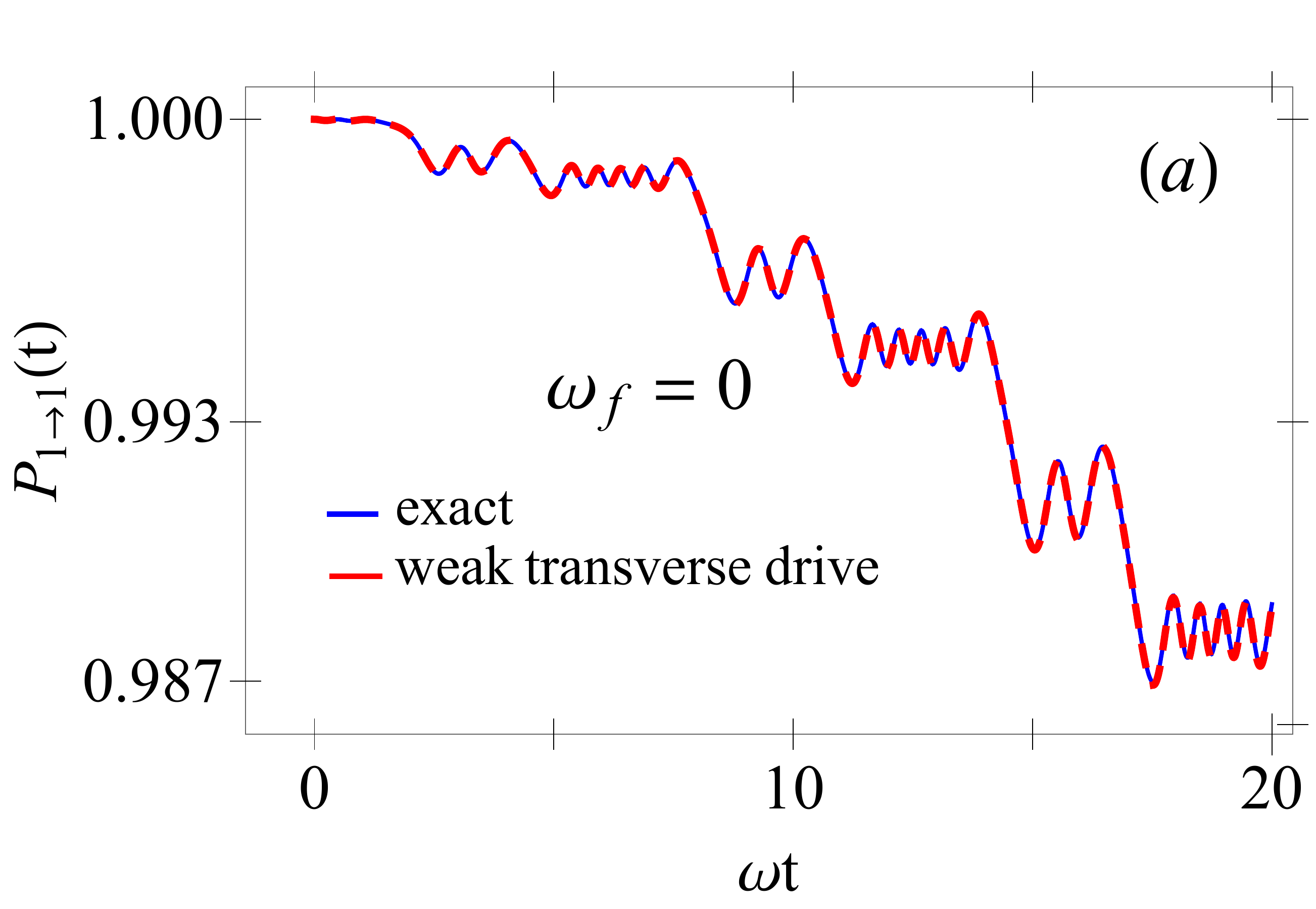}
\includegraphics[width=4.2cm, height=3.5cm]{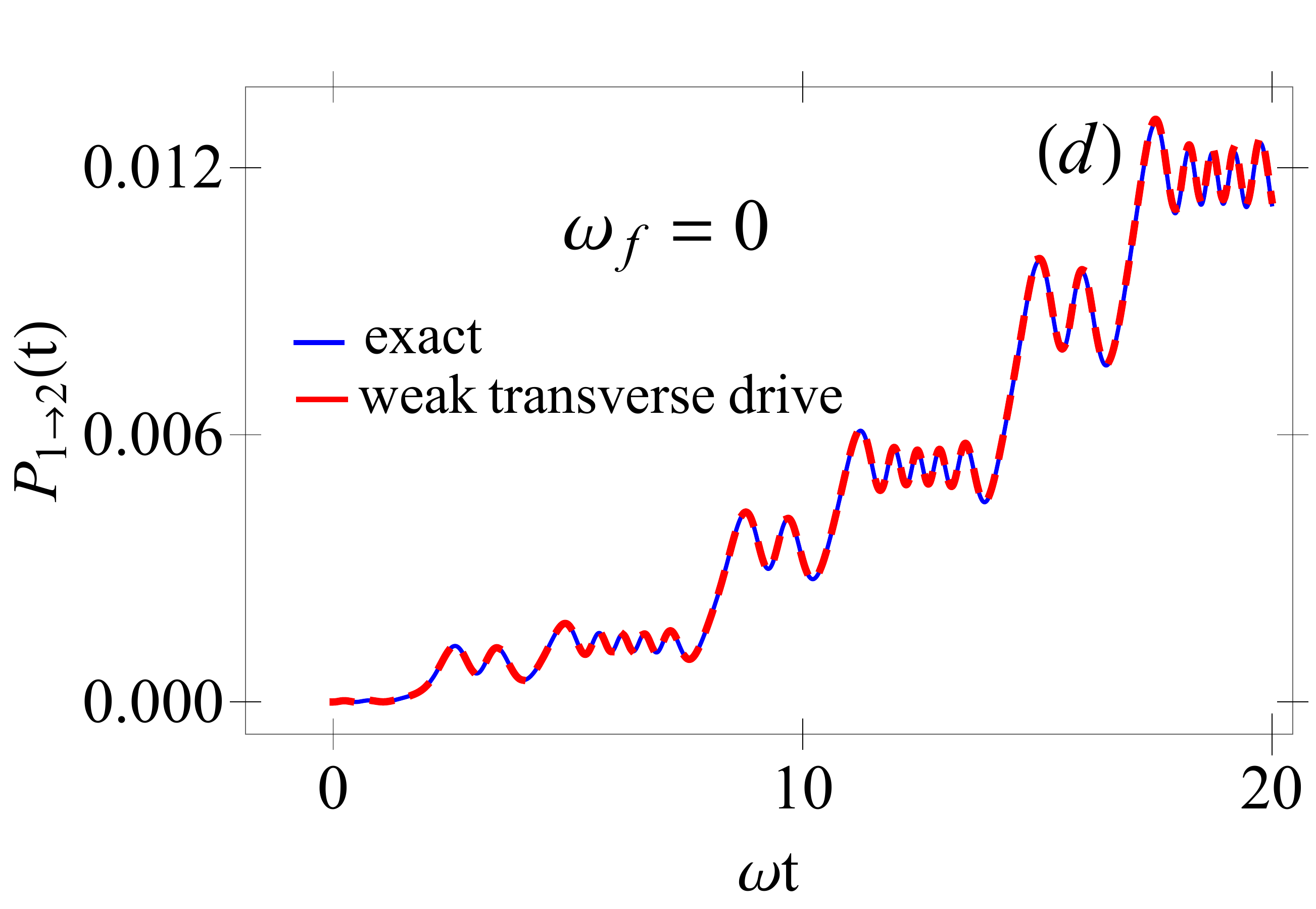}\\\vspace{-0.25cm}
\includegraphics[width=4.2cm, height=3.5cm]{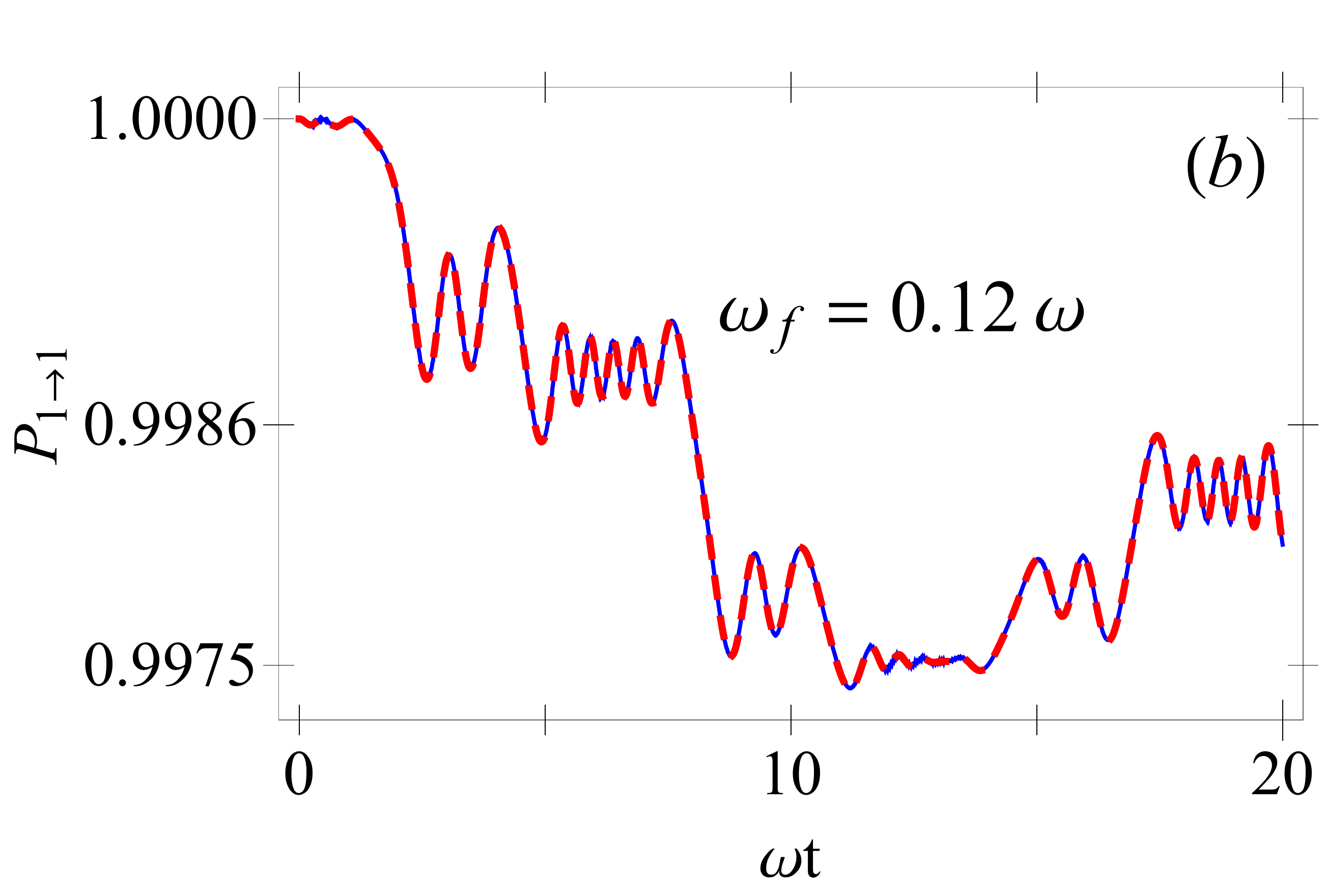}
\includegraphics[width=4.2cm, height=3.5cm]{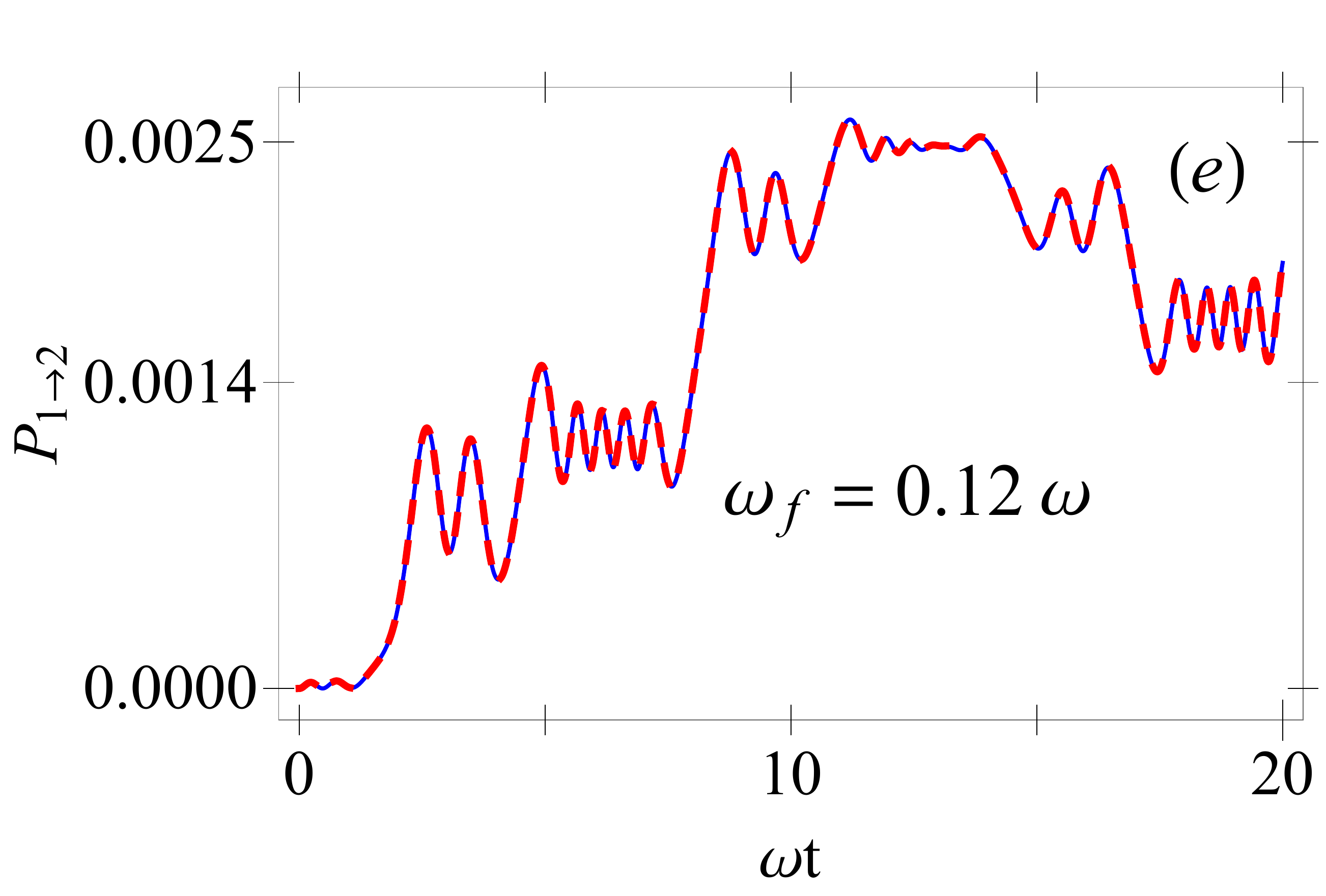}\\\vspace{-0.25cm}
\includegraphics[width=4.2cm, height=3.5cm]{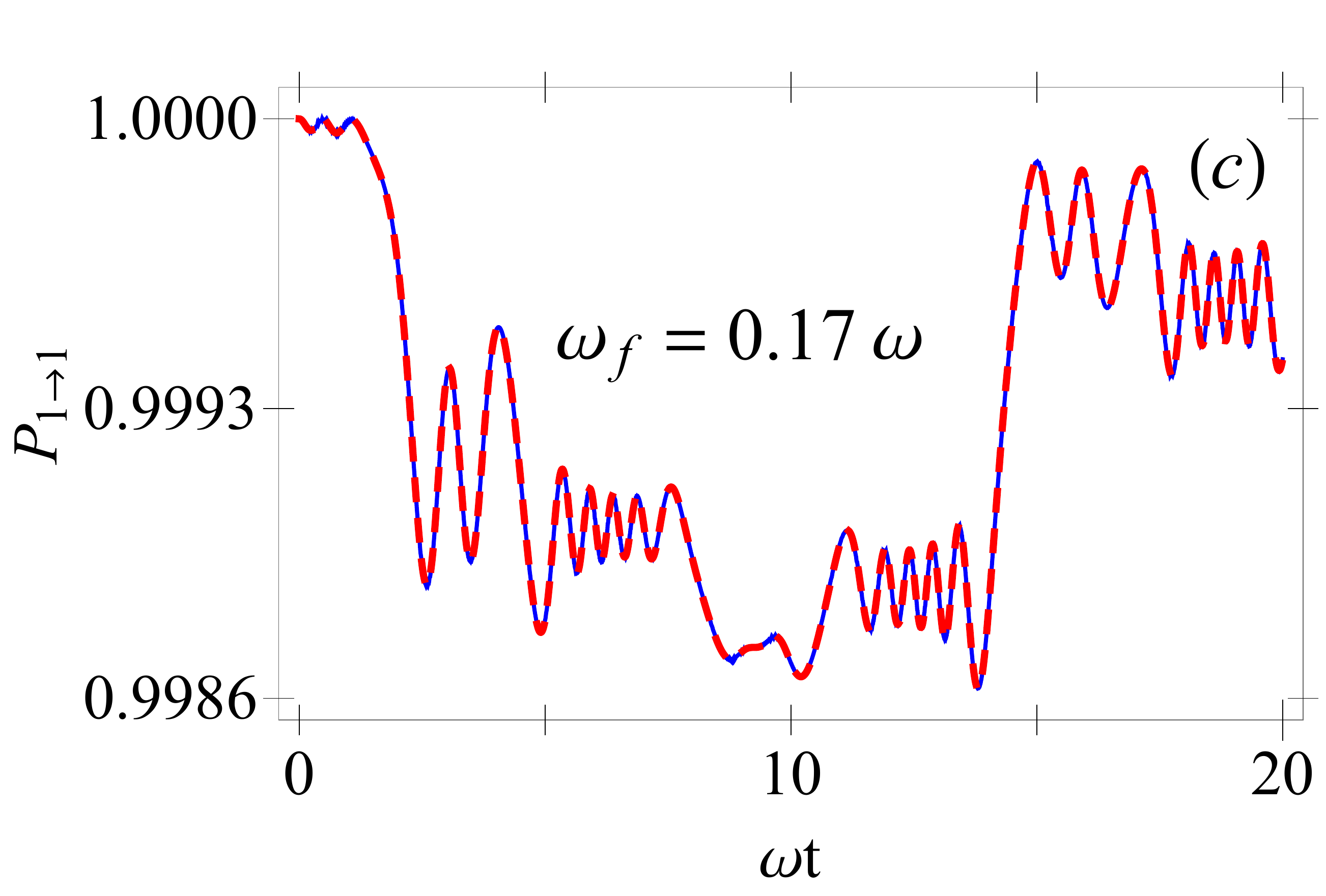}
\includegraphics[width=4.2cm, height=3.5cm]{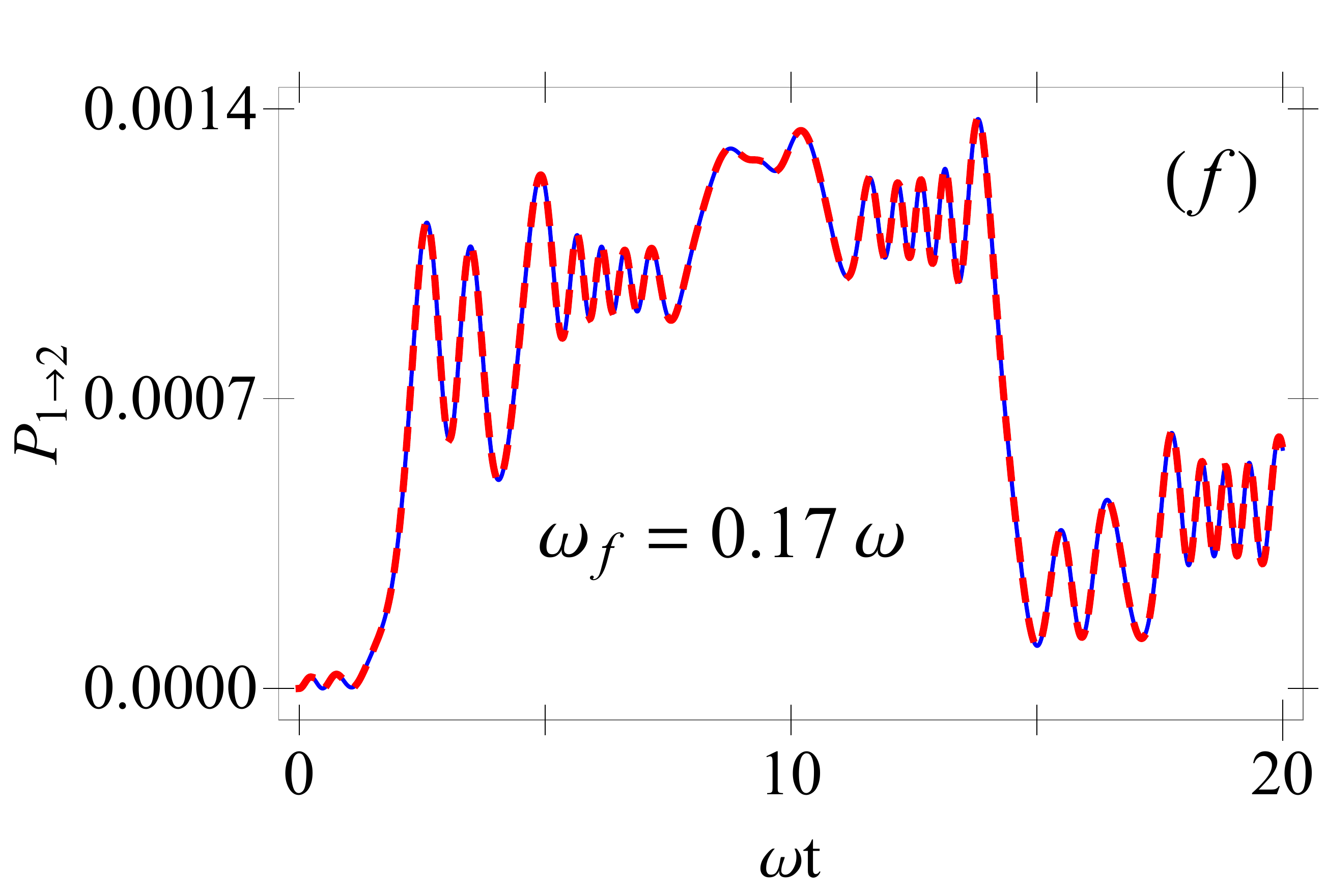}
\end{center}
\vspace{-0.75cm}
\caption{(Color Online) Populations $P_{1\to1}(\tau)$ (left panels)  and $P_{1\to2}(\tau)$ (right panels) obtained by numerically solving the TDSE (\ref{equ3}) with the model (\ref{equ4}) in the weak transverse drive limit and simultaneously displayed with data of the analytical formula in Eq.(\ref{equ11a}). Blue (solid)  lines indicate exact numerical results  while red (dashed) lines are analytical results. For implementation, we have used $A/\omega=10.5$, $\mathcal{A}_{f}/\omega=0.05$ and $D/\omega=3$. The infinite series in Eqs.(\ref{equ11c}) is truncated such that the index $n$ runs from $-20$ to $20$ and  the series converge towards the exact solution.} \label{Figure3}
\end{figure}

\begin{figure}[!h]
\vspace{-0.5cm}
\begin{center}
\includegraphics[width=7.5cm, height=6.5cm]{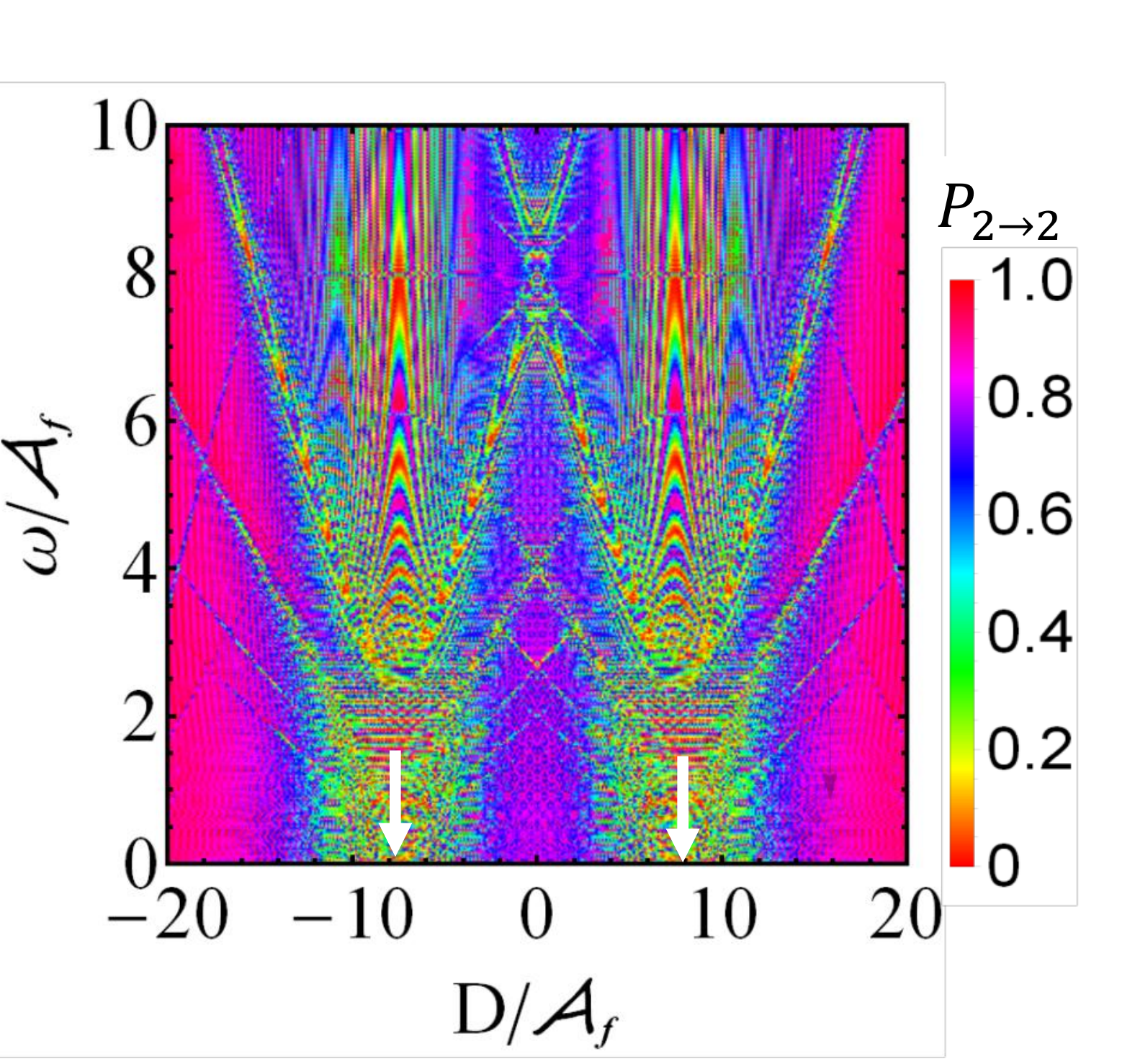}
\end{center}
\vspace{-0.75cm}
\caption{(Color Online)  Probability $P_{2\to2}(t)$ viewed as a function of $D/\mathcal{A}_{f}$ and $\omega/\mathcal{A}_{f}$ exactly calculated by numerically solving the TDSE (\ref{equ3}) for $A/\mathcal{A}_{f}=2.5$, $\omega_{f}/\mathcal{A}_{f}=8$ from the initial time $t_{0}=0$ to $t=50/\omega$. Interference patterns observed in Fig.\ref{Figure2bis} for weak $\omega_{f}$ appears here twice as $\omega_{f}$ is large. White arrows placed at $D=+\omega_{f}$ and $D=-\omega_{f}$ (the  frequency of the transverse drive is tuned to match the easy-axis anisotropy) correspond to the solutions of resonance conditions Eq.(\ref{equ11g}) when $n=0$. Given that $\omega_{f}$ is non-negative and that $D$ is naturally established for specific crystal lattices, we conclude that such a ''double'' interference patterns can only be observed in systems with large $D$ such as NVC in diamond where $D=2\pi\times2.88$GHz (see discussion in section \ref{Sec4}). This is a  manifestation of the $SU(3)$ dynamics. Indeed, if the $SU(3)$ symmetry is broken down by the substitution $D(S^{z})^{2}\to D'S^{z}$ in the Hamiltonian, the resonance equations (\ref{equ11g}) have a unique solution when $n=0$ namely $D'=\omega_{f}$ and these types of interference patterns are observed only once as in figure \ref{Figure2bis}.} \label{Fig4}
\end{figure}

As far as Eq.(\ref{equ5}) is concerned, we have to stress that it contains fast oscillating terms that induce undesired divergences into the dynamics of the system and complicate our task. The point of primary interest is then to remove all time-dependence from the longitudinal drive. We rotate the system from the Schr\"odinger to Dirac/interaction picture with the help of the gauge transformation $\mathbf{C}(t)=U(t)\phi(t)$. Here, $U(t)$ (population preserving) is a unitary operator $(U^{\dagger}(t)U(t)=U(t)U^{\dagger}(t)=\hat{\mathbf{1}})$ ensuring the rotation ($\hat{\mathbf{1}}$ being the $3\times3$ unit matrix, the symbol $\dagger$ indicates the Hermitian conjugate) and  $\phi(t)=[\phi_{1}(t), \phi_{2}(t), \phi_{3}(t)]^{T}$ a  three-component vector probability amplitude. $U(t)$ propagates the states $\phi(t)$ of the system in the $\phi$-basis from the past  instant $t_{0}$ when the driving fields are turned on, to an arbitrary time $t\ge t_{0}$ and removes all Abelian phase dynamics. The $\phi$-basis is directed by: 
\begin{equation} \label{equ6} 
i\frac{d\phi (t)}{dt}=\mathcal{H}_{\phi}(t)\phi(t). 
\end{equation} 
Here, 
\begin{equation} \label{equ7} 
\mathcal{H}_{\phi} (t)=U^{\dagger} (t)\mathcal{H}(t)U(t)-iU^{\dagger} (t)\frac{dU(t)}{dt} 
\end{equation} 
is nothing but $\mathcal{H}(t)$ in Eq.(\ref{equ1}) written in the $\phi$-basis.  The propagator $U(t)$ must be selected such that it sets all diabatic states on-resonance (zero detuning) i.e. the Hamiltonian $\mathcal{H}_{\phi} (t)$ becomes off-diagonal with zeros on the main diagonal. Thus, the unique rotation operator which satisfies this requirement reads:
\begin{equation}\label{equ8} 
U(t)=\exp\Big[-i\int_{t_{0}}^{t}dt_{1}[A\cos(\omega t_{1})S^{z}+D(S^{z})^{2}]\Big].
\end{equation}
In order to see the effective action of $U(t)$, let us introduce the dimensionless time $\tau=\omega t$ and importantly, the Jacobi-Anger relation\cite{Erderly} $e^{ix\sin(\tau)}=\sum_{n=-\infty}^{\infty}J_{n}(x)e^{in\tau}$ (where $J_{n}(x)$ is the Bessel function of the first kind  of order $n$ and argument $x$). Thus, the operator $U(\tau)$ acts as rotation onto the Hilbert space $\mathscr{H}$ and transfers the problem (\ref{equ6}) into the following: 
\begin{eqnarray}\label{equ9}
i\frac{d\phi(\tau)}{d\tau}=
\left[
{\begin{array}{*{20}c}
0 && \Omega_{-}(\tau) && 0\\
\Omega_{-}^{*}(\tau) && 0 && \Omega_{+}(\tau)\\
0  && \Omega_{+}^{*}(\tau) && 0
\end{array} } \right]\phi(\tau),
\end{eqnarray}
where 
\begin{eqnarray}\label{equ10}
\Omega_{\pm}(\tau)=\sum_{n=-\infty}^{\infty}\mathcal{J}_{n}^{\textmd{eff}}\Big(\frac{A}{\omega}\Big)\mathcal{K}_{n}^{\pm}(\tau),
\end{eqnarray}
with
\begin{eqnarray}\label{equ10aa}
\nonumber\mathcal{K}_{n}^{\pm}(\tau)=\exp\Big[i\Big(n\mp\frac{D}{\omega}\mp\frac{\omega_{f}}{\omega}\Big)\tau\Big]+\exp\Big[i\Big(n\mp\frac{D}{\omega}\pm\frac{\omega_{f}}{\omega}\Big)\tau\Big], \\
\end{eqnarray}
which are field-induced Rabi frequencies that measure the strength of couplings between diabatic states. The symbol $*$ in Eq.(\ref{equ9}) stands for the complex conjugate. We have defined the {\it effective} transverse field amplitude (effective Rabi frequency) caused by $ac$ fields as
\begin{eqnarray}\label{equ11}
\mathcal{J}_{n}^{\textmd{eff}}\Big(\frac{A}{\omega}\Big)=\frac{\mathcal{A}_{f}}{2\omega\sqrt{2}}J_{n}\Big(\frac{A}{\omega}\Big).
\end{eqnarray}
Note that no rotating-wave-approximation has been made\cite{comment5}. Let's remark that the time-evolution described by Eq.(\ref{equ9}) shows a sequence of consecutive  LZSM oscillations when the amplitude of the longitudinal driving field widely exceeds the uniaxis anisotropy ($A\gg D$) and the frequency of the transverse drive is weak enough.   Roughly speaking, the signature of LZSM oscillations in the population of levels is well pronounced in the extreme limit $\omega_{f}=0$ and  in the weak transverse drive limit ($\mathcal{A}_{f}\ll\omega$) as discussed in this subsection (figure \ref{Figure3}).  In order to understand such transitions, we use the arguments  defended in  Ref.\onlinecite{Kayanuma2000}. Indeed, the spectral decomposition of the probability amplitudes is $\phi_{\kappa}(t)=\sum_{r=-\infty}^{\infty}\Phi_{\kappa,r}(\omega)e^{ir\omega t}$ and the relevant eigen-spectrum reveals that the periodic modulation of control parameters in a ThLS, splits the states $\phi_{\kappa}(t)$ into an infinite series of sublevels with photon energy $r\omega$ (in the frequency unit as $\hbar=1$) such that any transition between $|\kappa'\rangle$ and $|\kappa\rangle$ is equivalent to successive LZSM transitions between sublevels. The corresponding LZSM interference patterns are reported in Fig.\ref{Figure2bis} for weak $\omega_{f}$ and in Fig.\ref{Fig4} for large $\omega_{f}$.

As for the case of TLSs, the presence of Bessel functions in the effective Rabi couplings (\ref{equ11})  indicates that there is a coherent destruction of tunneling (CDT) when the longitudinal driving field is turned such that the ratio $A/\omega$ achieves the zeroes of the Bessel function\cite{Hanggi}. When such a condition is realized, the Rabi frequencies $\Omega_{\pm}(\tau)$ in Eq.(\ref{equ10}) vanish and population transfers are inhibited. The ThLS remains in its original diabatic state (population return). In this case, the interference is destructive. In the opposite case when the system  completely goes to the excited-states, interferences between paths are constructive.  

It is relevant to derive the above equations because of their paramount importance in our discussion. They are quoted several times in the paper, not only when discussing the transverse drive limits, but also in the longitudinal drive limits. Let us now make use of them by returning to the main purpose of this subsection: the weak transverse drive limit. In this limit, $\mathcal{A}_{f}/\omega\ll1$, the effective Rabi interaction Eq.(\ref{equ11}), is weakened and $\mathcal{J}_{n}^{\textmd{eff}}(A/\omega)\ll1$, Eq.(\ref{equ9}) is readily integrated by perturbation between the instant $\tau_{0}=0$ and $\tau$.  The terms of the order $(\mathcal{A}_{f}/\omega)^{2}$ in the perturbation are neglected and the results read
\begin{eqnarray} \label{equ11a} 
\nonumber\mathbf{P}_{\kappa\to \kappa'}(\tau)\approx \left[\begin{array}{ccc} {1-p_{-}}(\tau) & {p_{-}}(\tau) & {0} \\ {p_{-}}(\tau) & {1-p_{+}(\tau)-p_{-}(\tau)} & {p_{+}}(\tau) \\ {0} & {p_{+}}(\tau) & {1-p_{+}}(\tau) \end{array}\right].\\
\end{eqnarray} 
Here, we have defined $p_{\pm}(\tau)=\tau^{2}F^{\pm}_{ss}(\tau)+\tau^{4}F^{\pm}_{cc}(\tau)$ with
$F^{\pm}_{ss}(\tau)=[G^{\pm}_{ss}(\tau)]^{2}$ and $F^{\pm}_{cc}(\tau)=[G^{\pm}_{cc}(\tau)]^{2}$
where
\begin{eqnarray}\label{equ11c}
 G^{\pm}_{ss}(\tau)=\sum_{n=-\infty}^{\infty}\mathcal{J}_{n}^{\textmd{eff}}\Big(\frac{A}{\omega}\Big)\Big[
j_{0}(\bar{\omega}_{\pm n}^{\pm}\tau)+j_{0}(\bar{\omega}_{\pm n}^{\mp}\tau)\Big], \quad
\end{eqnarray}
\begin{eqnarray}\label{equ11d}
\nonumber G^{\pm}_{cc}(\tau)=\sum_{n=-\infty}^{\infty}\mathcal{J}_{n}^{\textmd{eff}}\Big(\frac{A}{\omega}\Big)\Big[\frac{\bar{\omega}_{\pm n}^{\pm}}{2}j_{0}^{2}\Big(\frac{\bar{\omega}_{\pm n}^{\pm}\tau}{2}\Big)+\frac{\bar{\omega}_{\pm n}^{\mp}}{2}j_{0}^{2}\Big(\frac{\bar{\omega}_{\pm n}^{\mp}\tau}{2}\Big)\Big],\\
\end{eqnarray}
with $\bar{\omega}_{n}^{\pm}=(D\pm\omega_{f}-n\omega)/\omega$ and $j_{0}(z)=\sin(z)/z$ is the spherical Bessel function of first kind\cite{Erderly}. The element in position ($\kappa',\kappa$) in the transion matrix Eq.(\ref{equ11a})  represents the probability $P_{\kappa'\to \kappa}(\tau)$.  

The range of validity of our results is probed by numerical tests (see Fig.\ref{Figure3}). Exact numerical solutions of Eq.(\ref{equ9}) are calculated in the weak transverse drive amplitude limit $\mathcal{A}_{f}/\omega\ll1$ and simultaneously  depicted on the figure \ref{Figure3} with the corresponding probabilities in Eq.(\ref{equ11a}). Our results quite well reproduce the gross temporal profile of the exact probabilities and hold for arbitrary $\tau$, $A$, $D$, $\omega_{f}$ while the condition $\mathcal{A}_{f}/\omega\ll1$ is verified. Figure \ref{Figure3}, however tells us that any variation of the transverse drive frequency  $\omega_{f}$ has a  qualitative drastic incidence on the  levels populations. Mainly, LZSM-like oscillations disappear as $\omega_{f}$ increases. The consequences on interference patterns are presented in the  figure \ref{Fig4}. We observe that the patterns presented in figure \ref{Figure2bis} for small $\omega_{f}$ are doubled for large $\omega_{f}$. Therefore, the transverse drive splits the  spectrum in Fig.\ref{Figure2bis} into two components that are observed in Fig.\ref{Fig4}. Relevantly, this is also a consequence of the specific orientation of the two signals with respect to the quantization axis (crystal principal axis):  one parallel (longitudinal drive) and the other perpendicular (transverse drive)\cite{Comment2}.

Remark the population $P_{1\to3}(\tau)$ on the level $|3\rangle$ is negligibly small at any time for an initialization of the system in the state $|1\rangle$. This is dramatically the same situation for $P_{3\to1}(\tau)$ which is also of the order of $(\mathcal{A}_{f}/\omega)^{2}$ and negligible. This recalls that in the considered regime, the transverse field is not strong enough to produce an inversion of population (adiabatic transfer) from $|1\rangle$ to $|3\rangle$ and that the states $|1\rangle$ and $|2\rangle$, $|2\rangle$ and $|3\rangle$ interfere destructively. The system non-adiabatically  returns to its original diabatic state.  Thus, the state $|3\rangle$ can be adiabatically eliminated by preparing the system in the intermediate state $|2\rangle$ and maintaining the transverse drive such that $A_{f}\ll \Delta E$ (energy difference between the states with $m_{S}=0$ and $m_{S}=-1$) and no transition $m_{S}=0\leftrightarrow m_{S}=-1$ between neighboring states $|2\rangle \leftrightarrow |3\rangle$ occurs. These actions rescale the zero-energy splitting to $D=0$. As a direct consequence, one can set $\dot{C}_{3}(t)=0$ in the TDSE (\ref{equ5}).  Upon eliminating $C_{3}(t)$ in the remaining equations and neglecting terms of the order $(\mathcal{A}_{f}/\omega)^{2}$, the three-state problem reduces to the two-state 
$
i\dot{\mathbf{c}}(t)=\Big(\frac{1}{2}
A\cos(\omega t)\boldsymbol{\mathrm{\sigma}}_{z}+\mathcal{A}_{f}\cos(\omega_{f} t)\boldsymbol{\mathrm{\sigma}}_{x}\Big)\mathbf{c}(t),
$
where $\mathbf{c}(t)=[c_{1}(t), c_{2}(t)]^{T}$ with $c_{1,2}(t)=C_{1,2}(t)\exp[-\frac{i}{2}\int_{0}^{t}(A\cos(\omega t')+D)dt']$ and $\boldsymbol{\mathrm{\sigma}}_{x, z}$ are Pauli matrices. The model Hamiltonian in the above equation is quantitatively equivalent to (\ref{equ2}) when the aforementioned conditions are satisfied (numerical tests are done but not shown here).  It describes a TLS subject to a longitudinal and a weak transverse signal.  As we have already seen, the population in the state $|3\rangle$ typically becomes negligible and the dynamics of the ThLS can be described by the two-level model. Therefore, the transverse field, in the weak transverse regime can  then be utilized to adiabatically eliminate one of the states with extremal spin projections ($m_{S}=-1$ or $m_{S}=+1$) allowing for reducing the ThLS to a TLS. Such a technique is used for NVCs in diamond with a static longitudinal magnetic field which lifts the degeneracy between the states with $m_{S}=\pm 1$ (see Refs.\onlinecite{ Du, Zhou,  comment1, Ansari, Fuchs, Wubs} and the Subsection.\ref{Sec4}). The weak transverse regime as described here can be technically exploited as an alternative mean for achieving the same purpose. Though it is preferential to deal with TLSs,  such a reduction method technically cost given that it drastically reduces the number of controllable parameters and restrict the number of possible initial preparations of the system. This systematical leads to a qualitative loss of information. Indeed, after the reduction operation, the system can only be prepared in the intermediate diabatic state $|2\rangle$, the $SU(3)$ symmetry breaks down ($D=0$). Interference patterns such as those reported in Fig.\ref{Figure2bis} are not observed. Therefore, the {\it adiabatic reduction procedure} through the weak transverse drive regime is useful for reducing ThLSs to TLs but fails in helping to prepare qutrit which, as compared to qubit, is more robust against environmental nuisances, is implemented at room temperature and encodes more information (see introductory part). It remains therefore of paramount importance to consider all individual three states of the system (without eliminating any of them) as a platform for further realizing qutrit.

Consider now the case when the amplitude of the longitudinal driving field is weak $A/\omega\ll1$. In this case, LZSM-like oscillations disappears when $A<D$. The main contribution in the series of Bessel functions in Eqs.(\ref{equ11c}) and (\ref{equ11d}) comes from the zeroth order term ($n=0$). Populations on levels are of the form of Eq.(\ref{equ11a}) with 
\begin{eqnarray}\label{equ11e}
 G^{\pm}_{ss}=\mathcal{J}_{0}^{\textmd{eff}}\Big(\frac{A}{\omega}\Big)\Big[
j_{0}(\bar{\omega}^{\pm}_{0}\tau)+j_{0}(\bar{\omega}^{\mp}_{0}\tau)\Big], \quad
\end{eqnarray}
\begin{eqnarray}\label{equ11f}
 G^{\pm}_{cc}=\mathcal{J}_{0}^{\textmd{eff}}\Big(\frac{A}{\omega}\Big)\Big[\frac{\bar{\omega}^{\pm}_{0}}{2}j_{0}^{2}\Big(\frac{\bar{\omega}^{\pm}_{0}\tau}{2}\Big)+\frac{\bar{\omega}^{\mp}_{0}}{2}j_{0}^{2}\Big(\frac{\bar{\omega}^{\mp}_{0}\tau}{2}\Big)\Big].\quad
\end{eqnarray}
If in addition $\omega_{f}=0$, then $ G^{\pm}_{ss}=2\mathcal{J}_{0}^{\textmd{eff}}(A/\omega)
j_{0}(\bar{\omega}_{0}\tau)$ and $G^{\pm}_{cc}=\bar{\omega}_{0}\mathcal{J}_{0}^{\textmd{eff}}(A/\omega) j_{0}^{2}(\bar{\omega}_{0}\tau/2)$ with $\bar{\omega}_{0}=D/\omega$.

Let us also consider the complementary limit $A/\omega\gg1$. In this case, the Bessel functions mainly contribute at resonance points $n$ where $\bar{\omega}_{n}^{\pm}=0$. All contributions out of these points are due to fast oscillating terms, in average they vanish and are disregarded. Thus, the resonance conditions read:
\begin{subeqnarray}\label{equ11g}
D+\omega_{f}-n\omega=0,\\\nonumber\\
D-\omega_{f}-n\omega=0.
\end{subeqnarray}
To understand these observations, one can notice that the limit $\omega\to0$ supports the assumption $A/\omega\gg1$ and that $\lim_{\omega\to0}\frac{j_{0}(z/\omega)}{\omega}=\delta(z)$ where $\delta(z)$ is the Dirac Delta function nonzero only at point $z=0$. Thus, applying this strategy, functions of the form $\delta(D\pm\omega_{f}-n\omega)$  appear, telling us that only the values of $n$ satisfying (\ref{equ11g}) contribute to the series of Bessel functions. As an immediate consequence, $j_{0}(\bar{\omega}_{\pm n}^{\pm}\tau)=1$, $G^{\pm}_{cc}=0$ and only 
\begin{eqnarray}\label{equ11h}
 G^{\pm}_{ss}=\Big[\mathcal{J}_{\pm(D\pm\omega_{f})/\omega}^{\textmd{eff}}\Big(\frac{A}{\omega}\Big)
+\mathcal{J}_{\pm(D\mp\omega_{f})/\omega}^{\textmd{eff}}\Big(\frac{A}{\omega}\Big)\Big], 
\end{eqnarray}
contribute to the transition probabilities. If now $A/\omega$ is exceedingly large, the Bessel function is asymptotically approached as\cite{Erderly}  
\begin{eqnarray}\label{equ11i}
J_{\rho}(z)\approx \sqrt{\frac{2}{\pi z}}\cos\Big(z-\frac{\pi \rho}{2}-\frac{\pi}{4}\Big),\quad z\gg1.
\end{eqnarray}
Populations on diabatic levels obey a periodic dependence  as
\begin{eqnarray}\label{equ11j}
 G^{\pm}_{ss}\approx\sqrt{\frac{4\mathcal{A}_{f}^{2}}{\pi A\omega}}\cos\Big(\frac{A}{\omega}\mp\frac{\pi D}{2\omega}-\frac{\pi}{4}\Big)\cos\Big(\frac{\pi\omega_{f}}{2\omega}\Big).
\end{eqnarray}
The action of the longitudinal field is canceled by that of the transverse field when the later is turned such that $\omega_{f}=(2N+1)\omega$. Indeed, the system is trapped and completely returns to its original diabatic state after interactions $P_{\kappa'\to\kappa'}(\tau)=1$ as $G^{\pm}_{ss}(\tau)=0$.  

\subsection{Strong transverse driving limit $\mathcal{A}_{f}\gg \omega$}\label{adiab}

This limit is somehow equivalent to that of adiabatic evolutions $A\omega\ll\mathcal{A}^{2}_{f}$ when the Rabi coupling (transverse drive) weakly depends on time. For this reason and for the sake of universality, we construct a general adiabatic theory which holds for arbitrary three-state Hamiltonians of the form (\ref{A0}) and apply the results of our investigations to the generic model (\ref{equ1}) in the strong transverse driving limit. It should be noted however that, for application of this theory, regardless of the Hermitivity of the Hamiltonian, two conditions should be satisfied: (i)  direct transitions between states with extremal spin projections are forbidden such that transitions between them can only be achieved through a middle channel; (ii) the middle diabatic state is on-resonance and serves as a shuttle between the lower and upper states (the model (\ref{equ1}) satisfies these requirements).   Thus, we move to the composite Hilbert space $\tilde{\mathscr{H}}$ generated by the time-dependent basis vectors $|\varphi_{1}(t)\rangle$, $|\varphi_{2}(t)\rangle$ and $|\varphi_{3}(t)\rangle$ that are eigenstates of the perturbed Hamiltonian (adiabatic states). They are  orthogonal $\langle \varphi_{\kappa'}(t)|\varphi_{\kappa}(t)\rangle=\delta_{\kappa\kappa'}$ and preserve their total norm $\sum_{\kappa=1}^{3}|\varphi_{\kappa}(t)\rangle\langle\varphi_{\kappa}(t)|=1$ (in the absence of dissipation) at any arbitrary time $t$ and satisfy the eigenvalue equation
$\mathcal{H}(t)|\varphi_{n}(t)\rangle=E_{n}(t)|\varphi_{n}(t)\rangle$ ($n,\kappa=1,2,3$ throughout this subsection). Here, $E_{n}(t)$ are eigen-energies given in compact form by the expression 
\begin{eqnarray}\label{a2}
E_{n}(t)=\frac{\omega_{+}(t)+\omega_{-}(t)}{3}+2\sqrt{\frac{p(t)}{3}}\cos\Big[\frac{\vartheta_{n}(t)}{3}\Big],
\end{eqnarray}
where
\begin{eqnarray}\label{a3}
\vartheta_{n}(t)=\arccos\Big[\frac{3q(t)}{2p(t)}\sqrt{\frac{3}{p(t)}}\Big]-\delta_{n},
\end{eqnarray}
with $\delta_{1}=4\pi$, $\delta_{2}=2\pi$ and $\delta_{3}=0$. $\omega_{\pm}(t)$ are detunings in the generalized Hamiltonian (\ref{A0}). The functions $p(t)$ and $q(t)$ are deferred in appendix \ref{App1} where they are given by Eqs.(\ref{A3}) and (\ref{A4}) respectively while the eigenstates associated with eigen-energies Eq.(\ref{a2}) are
\begin{eqnarray} \label{a4} 
\nonumber {\left| \varphi _{n} (t) \right\rangle} =\frac{1}{\mathcal{N}_{n} (t)}\sum_{\ell=1}^{3}f_{n\ell} (t){\left| \ell \right\rangle}, \quad\mathcal{N}_{n}(t)=\sqrt{\sum_{\ell=1}^{3}f_{n\ell}^{2}(t)}.\\
\end{eqnarray} 
For all $n$, the functions $\mathcal{N}_{n}(t)$ are normalization factors whereas the functions $f_{n\ell} (t)$ are written in explicit form in Eqs.(\ref{A2a})-(\ref{A2c}). This basis is convenient to reveal the wave-function during adiabatic stages.   It is equally suited when the Hamiltonian is constant in time and describes three-level atoms undergoing Rabi oscillations\cite{Ansari}. This is precisely another purpose of this subsection. Indeed, some of the approximations made later lead to constant Hamiltonians and the relevant TDSEs need to be solved. As we already know, a system evolves adiabatically when the corresponding Hamiltonian is quasi-constant in time. To face all these problems simultaneously, we introduce the following  analytical scheme. Let $|\Psi (t)\rangle$  expressed in $\tilde{\mathscr{H}}$ be 
\begin{equation} \label{a5} 
|\Psi (t)\rangle=\sum_{n=1}^{3}a_{n}(t)|\varphi_{n}(t)\rangle,
\end{equation}
where $a_{n}(t)=\langle\varphi_{n}(t)|\Psi(t)\rangle\in\mathbb{C}$ (coefficients of the expansion) are probability amplitudes in the adiabatic basis or the norm of $|\Psi(t)\rangle$ along the direction of $|\varphi_{n}(t)\rangle$ in the same basis.  Thus, $|\varphi_{n}(t)\rangle$ can readily be decomposed in the fixed basis of $\mathscr{H}$ (diabatic basis) as
\begin{equation} \label{a6} 
|\varphi_{n}(t)\rangle=\sum_{\kappa=1}^{3}w_{\kappa n}(t)|\kappa\rangle,
\end{equation}
where $w_{n\kappa}(t)=\langle \kappa|\varphi_{n}(t)\rangle=f_{n\kappa}(t)/\mathcal{N}_{n}$ are projections of $|\varphi_{n}(t)\rangle$ onto the direction of $|\kappa\rangle$ in the diabatic basis. They obey the properties
\begin{equation} \label{aa6} 
\sum_{\kappa=1}^{3}w_{\kappa i}(t)w_{\kappa j}(t)=\delta_{ij}, \quad {\rm and}\quad \sum_{\ell,\kappa=1}^{3}w_{\ell \kappa}^{2}(t)=3,
\end{equation}
where $3$ in the right hand side of the second property reminds us of the dimension of the Hilbert space. Given a single crossing point  $t_{cr}$, far from the right and left of $t_{cr}$, non-adiabatic and adiabatic evolutions follow the same trajectories (one can also see figure \ref{Figure0}). Thus, if the system is initially prepared at time $t_{0}$ far from the left of $t_{cr}$ in the diabatic state $|\kappa'\rangle$, it is slowly transported by the state $|\varphi_{n}(t)\rangle$ to a different diabatic state $|\kappa\rangle\neq|\kappa'\rangle$ at final time $t\ge t_{0}$. Then, $|\varphi_{n}(t\ll t_{cr})\rangle=|\kappa'\rangle$ and $|\varphi_{n}(t\gg t_{cr})\rangle=|\kappa\rangle$ thus, $w_{n\kappa}(t\ll t_{cr})=\delta_{\kappa'\kappa}$ and $w_{n\kappa'}(t\gg t_{cr})=\delta_{\kappa'\kappa}$. Adiabatic states $|\varphi_{1,3}(t)\rangle$ and $|\varphi_{2}(t)\rangle$  act then as shuttles, mediating population transport between diabatic states. This process takes a relatively long running time,  causing a loss of coherence and spontaneous emission when the system is open to its environment. 

The matrix elements $w_{n\kappa}(t)$ allow us to rotate the system from diabatic to adiabatic basis. By substituting (\ref{a6}) into (\ref{a5}) and comparing the result with (\ref{equ3}), one finds that diabatic and adiabatic probability amplitudes are related as 
\begin{equation} \label{a7} 
C_{\kappa}(t)=\sum_{n=1}^{3}w_{\kappa n}(t)a_{n}(t).
\end{equation}
Then, by inserting Eq.(\ref{a7}) into the TDSE (\ref{equ5}), one shows that adiabatic probability amplitudes subjected to the initial condition $a_{\kappa'}(t_{0})=\delta_{\kappa'\kappa}$ obey the linear differential equation
\begin{equation} \label{equ4d} 
i\frac{d}{dt}a_{n}(t)=E_{n}(t)a_{n}(t)-i\sum_{\kappa=1}^{3}\nu_{n\kappa}(t)a_{\kappa}(t),
\end{equation}
where the $\nu_{n\kappa}(t)=\langle \varphi_{n}(t)|\dot{\varphi}_{\kappa}(t)\rangle=-\nu_{\kappa n}^{\dagger}(t)$ estimate the strength of non-adiabatic couplings between adiabatic states. In the same diabatic state, $\nu_{nn}(t)=0$ and for non-degenerate states
\begin{equation} \label{equ4dd} 
\nu_{n\kappa}(t)=-\frac{\langle \varphi_{n}(t)|(\partial_{t}\mathcal{H})|\varphi_{\kappa}(t)\rangle}{E_{n}(t)-E_{\kappa}(t)},\quad n\neq\kappa.
\end{equation}
Note that $\nu_{n\kappa}(t)$ vanish for constant-in-time Hamiltonians.  Equation (\ref{equ4d}) is exact (no approximation has been made) and is purely equivalent to (\ref{equ4}) written in a different basis.  However, as the Hamiltonian slowly varies in time, adiabatic evolution requires adiabatic states to not talk at all. In other words, non-adiabatic couplings between adiabatic states should be less than energy splitting between them ($\nu_{n\kappa}(t)\ll |E_{n}(t)-E_{\kappa}(t)|$ when $n\neq\kappa$). For strong adiabatic evolution as considered here,  non-adiabatic couplings  are completely eliminated. Then, the general condition for adiabatic evolution reads
\begin{equation} \label{equ4e} 
\nu_{n\kappa}(t)=0, \quad n\neq\kappa.
\end{equation}
This condition ensures that the eigenstates of the system are separated by large gaps such that no transition between them occurs. The system mainly remains in the same adiabatic state (eigenstate), consequently changing its diabatic state. Adiabatic states then realize population inversion. The condition (\ref{equ4e})  diagonalizes the Hamiltonian and is trivial (automatically holds) for Rabi-like Hamiltonians as we just pointed out. Equation (\ref{equ4d}) is readily integrated.  Assuming that the system starts off at time $t_{0}$ in the diabatic states $|\kappa'\rangle$ and ends up at arbitrary time $t$ in a different diabatic state $|\kappa\rangle$,  one finds that the total wave function during strong adiabatic evolutions reads
\begin{equation} \label{equ4f} 
|\Psi(t)\rangle=\sum_{n,\kappa=1}^{3}w_{\kappa n}(t)e^{-i\Lambda_{n}(t,t_{0})}w_{\kappa'n}(t_{0})|\kappa\rangle,
\end{equation}
where
\begin{equation} \label{equ4gg} 
\Lambda_{n}(t,t_{0})=\int_{t_{0}}^{t}E_{n}(t')dt',
\end{equation}
is the surface covered/swept by the eigen vector $|\varphi_{n}(t)\rangle$ within the time interval $]t_{0},t]$. It corresponds to the dynamical phase acquired by the system during adiabatic evolutions. Let us note that the geometric Berry's phase\cite{Berry}
$ 
\gamma_{n}(t,t_{0})=i\int_{t_{0}}^{t}\langle \varphi_{n}(t')|\frac{d}{dt'}|\varphi_{n}(t')\rangle dt' 
$ 
 does not contribute to the transition in this approximation given that $\nu_{nn}(t)=0$. As already pointed out, some of our approximations open onto constant models. We will call equation (\ref{equ4f}) when this situation is encountered. As a general algorithm for solving any three-level problem of the form (\ref{A0}), one first has to evaluate the eigenvalues and eigenvectors of the model. This helps to compute the rotation matrix $\mathbf{W}(t)$ and consequently $|\Psi(t)\rangle$. The probability for the system to be transferred from the diabatic state $|\kappa'\rangle$ to $|\kappa\rangle$ from an arbitrary time $t_{0}$ to $t\ge t_{0}$ is obtained from (\ref{equ4f}) by projecting the vector $|\Psi(t)\rangle$ onto the direction $|\kappa\rangle$ in the Hilbert space $\mathscr{H}$. Thus, in compact form,  $P_{\kappa'\to \kappa}(t_{0},t)=|\langle \kappa|\Psi(t)\rangle|^{2}$ reads:
\begin{widetext}
\begin{eqnarray}\label{equ4h}
\nonumber P_{\kappa'\to \kappa}(t_{0},t)=\Big[w_{\kappa'1}(t_{0})w_{\kappa 1}(t)+w_{\kappa'2}(t_{0})w_{\kappa2}(t)\cos[\phi_{ad}(t_{0},t)]+w_{\kappa'3}(t_{0})w_{\kappa 3}(t)\cos[\vartheta_{ad}(t_{0},t)]\Big]^{2}\\+
\Big[w_{\kappa'2}(t_{0})w_{\kappa 2}(t)\sin[\phi_{ad}(t_{0},t)]+w_{\kappa'3}(t_{0})w_{\kappa3}(t)\sin[\vartheta_{ad}(t_{0},t)]\Big]^{2}.
\end{eqnarray}
\end{widetext}
Here, $\phi_{ad}(t_{0},t)=\Lambda_{1}(t_{0},t)-\Lambda_{2}(t_{0},t)$ and $\vartheta_{ad}(t_{0},t)=\Lambda_{1}(t_{0},t)-\Lambda_{3}(t_{0},t)$ are respectively the surfaces in between the curves of $E_{1}(t)$ and $E_{2}(t)$ on one hand, that of $E_{1}(t)$ and $E_{3}(t)$ on the other hand. Expression (\ref{equ4h}) holds for any ThLS described  by a slowly-varying-in-time Hamiltonian of the form (\ref{A0}) with the hidden/trivial dynamical symmetry of the $SU(3)$ group or not.  

As yet another important aspect of our result, we should mention that it takes into account all initial moments where the system may be prepared. This gives a wide range of possibilities to adiabatically manipulate a ThLS satisfying the conditions (i) and (ii) and whose Hamiltonian slowly changes in time or in an extreme limit, does not vary in time at all (time-independent Hamiltonian). 
\begin{figure}[!h]
\vspace{-0.5cm}
\begin{center}
 \includegraphics[width=4cm, height=4cm]{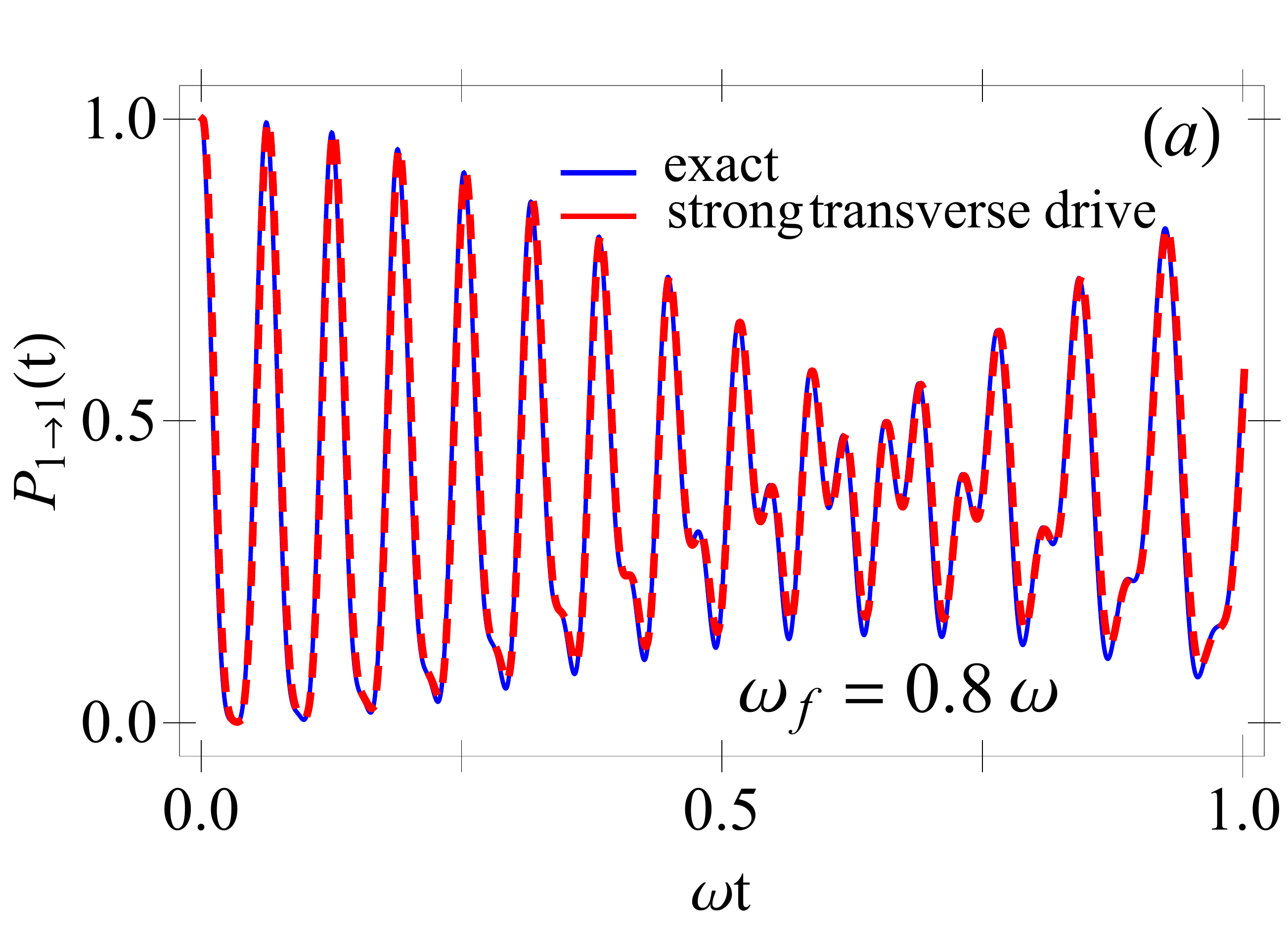}
\includegraphics[width=4cm, height=4cm]{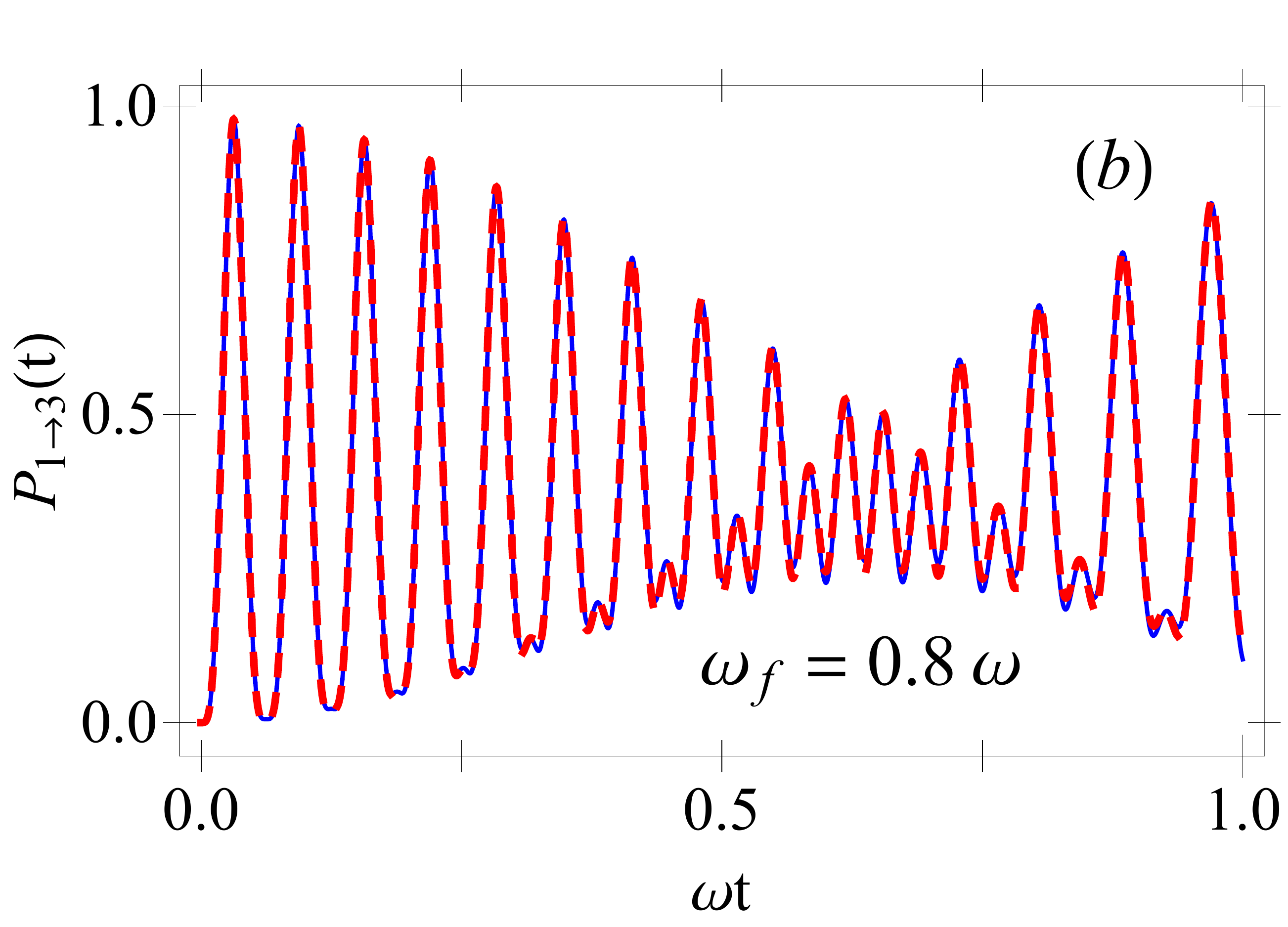}\\\vspace{-0.25cm}
\includegraphics[width=4cm, height=4cm]{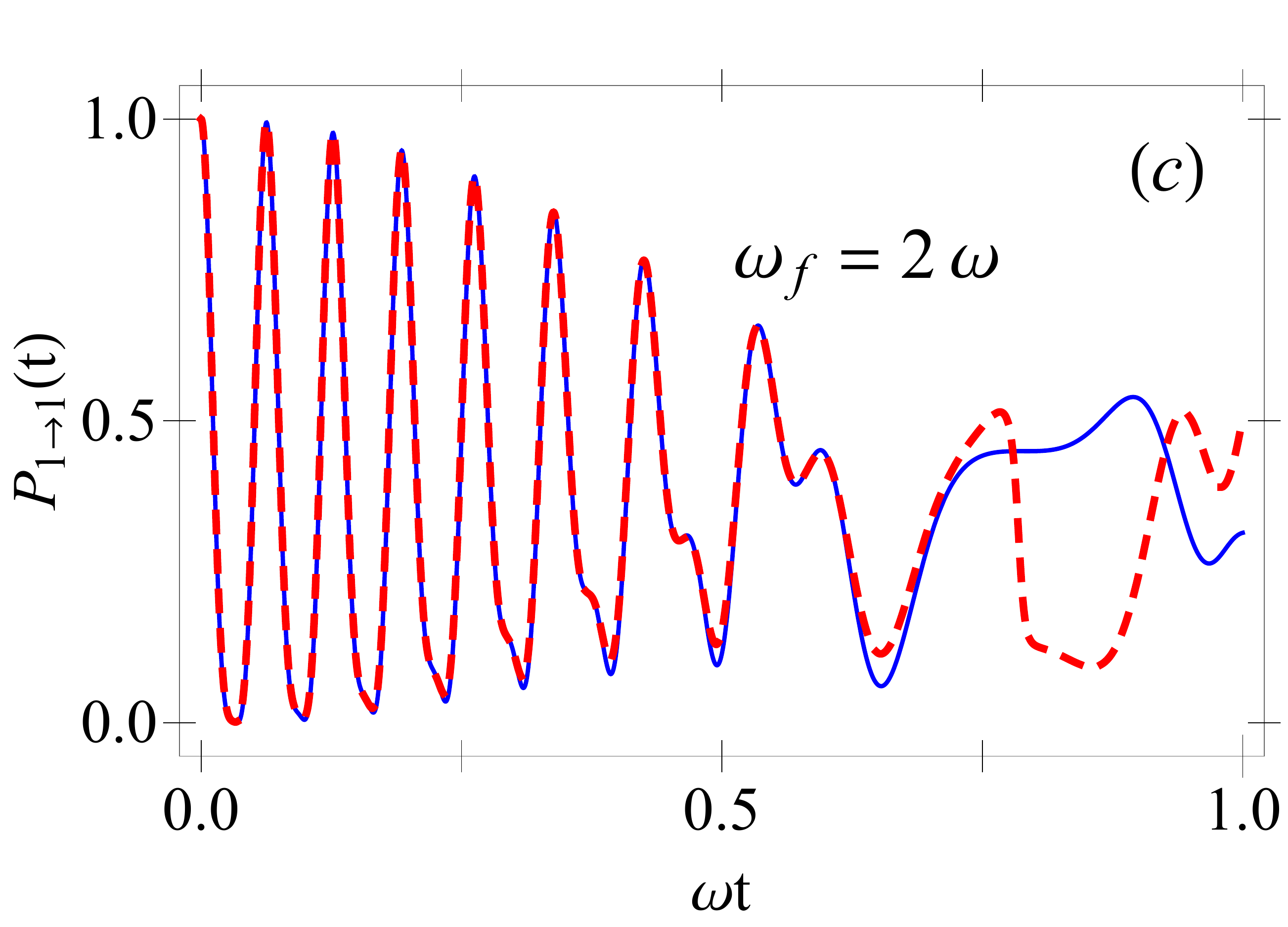}
\includegraphics[width=4cm, height=4cm]{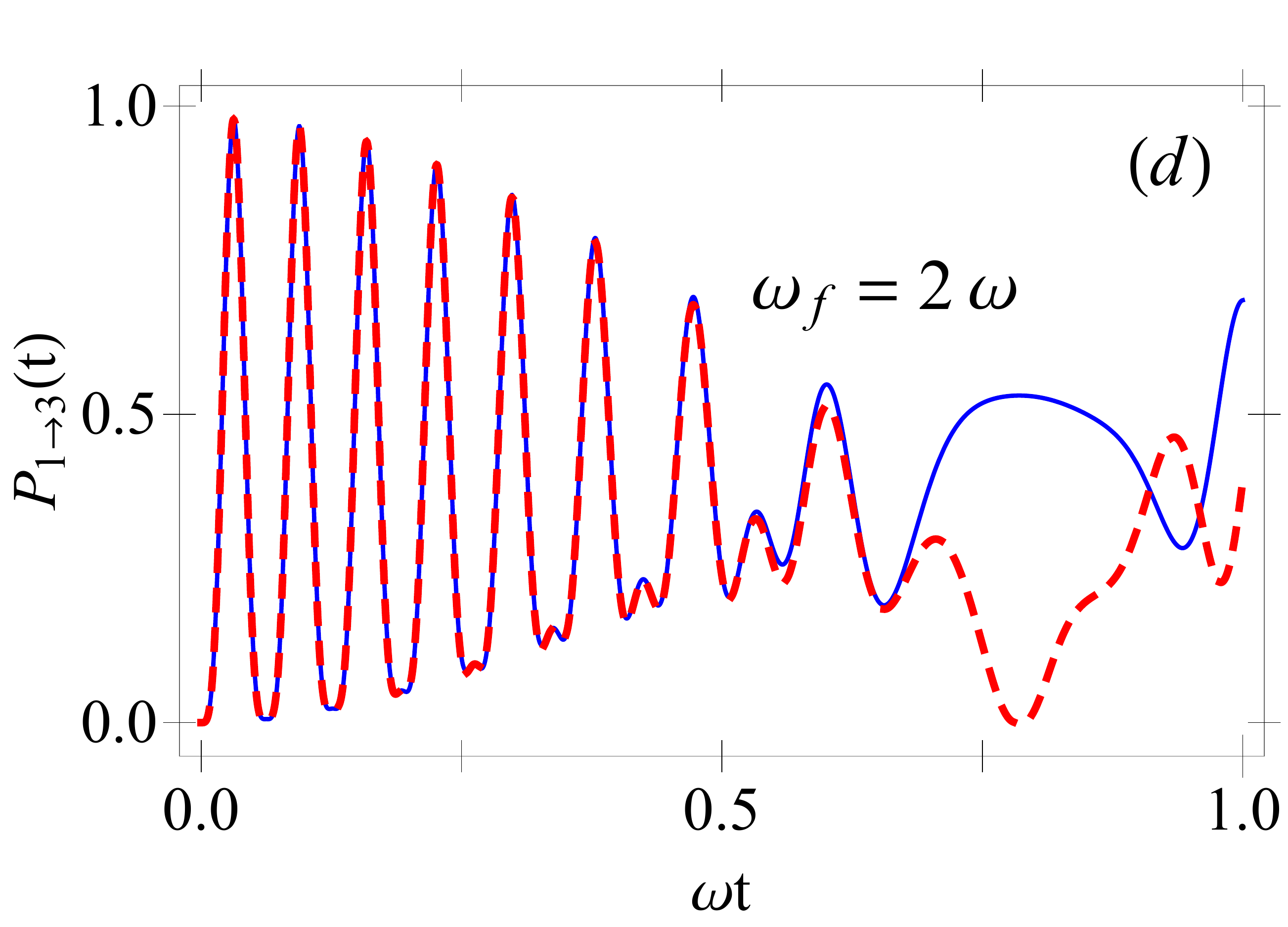}
\end{center}
\vspace{-0.75cm}
\caption{ (Color Online) Comparison between numerical solutions of Eq.(\ref{equ3}) with the model (\ref{equ4}) and transition probabilities $P_{\kappa'\to\kappa}(\tau)$ calculated from Eqs.(\ref{equ4h}). To calculate all graphs we have used $A/\omega=10$ (strong longitudinal amplitude drive), $\mathcal{A}_{f}/\omega=100$ and $D/\omega=5$. It appears from here that Eqs.(\ref{equ4h}) is useful to probe ThLSs in the regime $\omega_{f}/\omega\ll1$ and $\mathcal{A}_{f}/\omega\gg1$. For an application in the regime $\omega_{f}/\omega\gg1$, two conditions should be satisfied: firstly $\omega t\ll1$ and secondly $\mathcal{A}_{f}/\omega$ exceedingly large.} \label{Figure6}
\end{figure}

Adiabatic evolutions of the
 ThLS driven back and forth through a crossing are here treated with the aid of the  result (\ref{equ4h}).
The model (\ref{equ1}) satisfies the condition (i) and (ii) and the result (\ref{equ4h}) applies. In order to confirm the validity of the formula for this specific case, a numerical test is implemented. Essentials of our results are displayed in Fig.\ref{Figure6}. The graphs therein are barely discernible, numerical and analytical data are in good agreement. We have however observed that when the transverse drive frequency $\omega_{f}$ increases ($\sim 10^{1}\omega$), our result holds only for exceedingly large values of the transverse drive amplitude $\mathcal{A}_{f}$ ($\sim 10^{3}\omega$). Otherwise, for weakly large values of $\mathcal{A}_{f}$ ($10\omega\le\mathcal{A}_{f}\le100\omega$),  our result is relevant only in a short time interval (much more smaller than the period of the transverse drive) after the switch-on of the transverse drive ($t\ll1/\omega_{f}$). This is an indication that our result is valid for arbitrary detuning when the Rabi coupling (transverse) weakly depends on the time, that is $\omega_{f} t\ll1$ (see Fig.\ref{Figure6}(c) and \ref{Figure6}(d)). The reason for this is that as $\omega_{f}$ increases, the transverse drive rapidly oscillates, the condition for adiabatic evolution (\ref{equ4e}) is no longer satisfied and our approximation collapses. Thus, to revive our solution in this regard such that the condition (\ref{equ4e}) becomes valid, one has to significantly increase $\mathcal{A}_{f}$ to attenuate the effects of large $\omega_{f}$.  However, we have to stress that this situation might be weakly relevant for experimental realizations. Indeed, we have seen that for $\omega_{f}=8\omega$, one must take $\mathcal{A}_{f}=1000\omega$ to apply our result. 

\section{Longitudinal driving approximations}\label{Sec2}

\subsection{Weak longitudinal driving limit $A\ll\omega$}\label{slow}

In the weak longitudinal driving limit, high-frequency fields (fast oscillating) average out and Eq.(\ref{equ9}) simplifies. Indeed, in the expansion of Bessel functions, we  consider only dominating terms (zero-order, $n=0$)  neglecting higher-order terms. As a direct consequence, the effective Rabi frequencies in Eq.(\ref{equ10}) becomes $\Omega_{+}=\Omega_{-}^{*}$ and in the resulting equations, we set $\tau=\omega t$ and apply a second gauge $\phi(\tau)=\exp[i(\frac{D}{\omega}(S^{z})^{2}+\frac{\omega_{f}}{\omega}S^{z})\tau]\mathbf{F}(\tau)$ where the three-component vector probability amplitude $\mathbf{F}(\tau)=[F_{1}(\tau),F_{2}(\tau),F_{3}(\tau)]^{T}$ obeys (where we have used a RWA)
 \begin{align}\label{equ16}
i\dot{\mathbf{F}}(\tau)=
\Big[
\omega_{0} S^{z}+\Omega_{0}S^{x}+\Delta_{0}(S^{z})^{2}\Big]\mathbf{F}(\tau),
\end{align}
where
\begin{eqnarray} 
\Omega_{0}=\frac{\mathcal{A}_{f}}{\omega}J_{0}\Big(\frac{A}{\omega}\Big),\quad \Delta_{0}=\frac{D}{\omega},\quad {\rm and}\quad \omega_{0}=\frac{\omega_{f}}{\omega}.
\end{eqnarray}
 When $A/\omega$ coincides with one of the zeros of the Bessel function $J_{0}(A/\omega)$, CDT occurs. Rabi interactions are considerably reduced by the factor $J_{0}(A/\omega)$. It should be relevant for further purposes to note that the problem in Eq.(\ref{equ16}) mimics the dynamics of a three-level atom undergoing Rabi oscillations. The Hamiltonian  can then be exactly diagonalized. This immediately invites the theory presented in Subsection \ref{adiab}.  The relevant Hamiltonian is of the form (\ref{A0}) with all identical constant-in-time Rabi interactions $\Delta_{ij}(\tau)=\Omega_{0}/\sqrt{2}$, $\Delta_{ii}(\tau)=0$, $(i,j)=1,2,3$ and detunings $\omega_{\pm}=\Delta_{0}\pm\omega_{0}$. Evolutions directed by such a Hamiltonian are adiabatic. The eigen-energies are given  by (\ref{a2}) and the eigenvectors by (\ref{a4}). Probabilities are then unquestionably given by (\ref{equ4h}).

\begin{figure}[]
\vspace{-0.5cm}
\begin{center}
 \includegraphics[width=4.2cm, height=4cm]{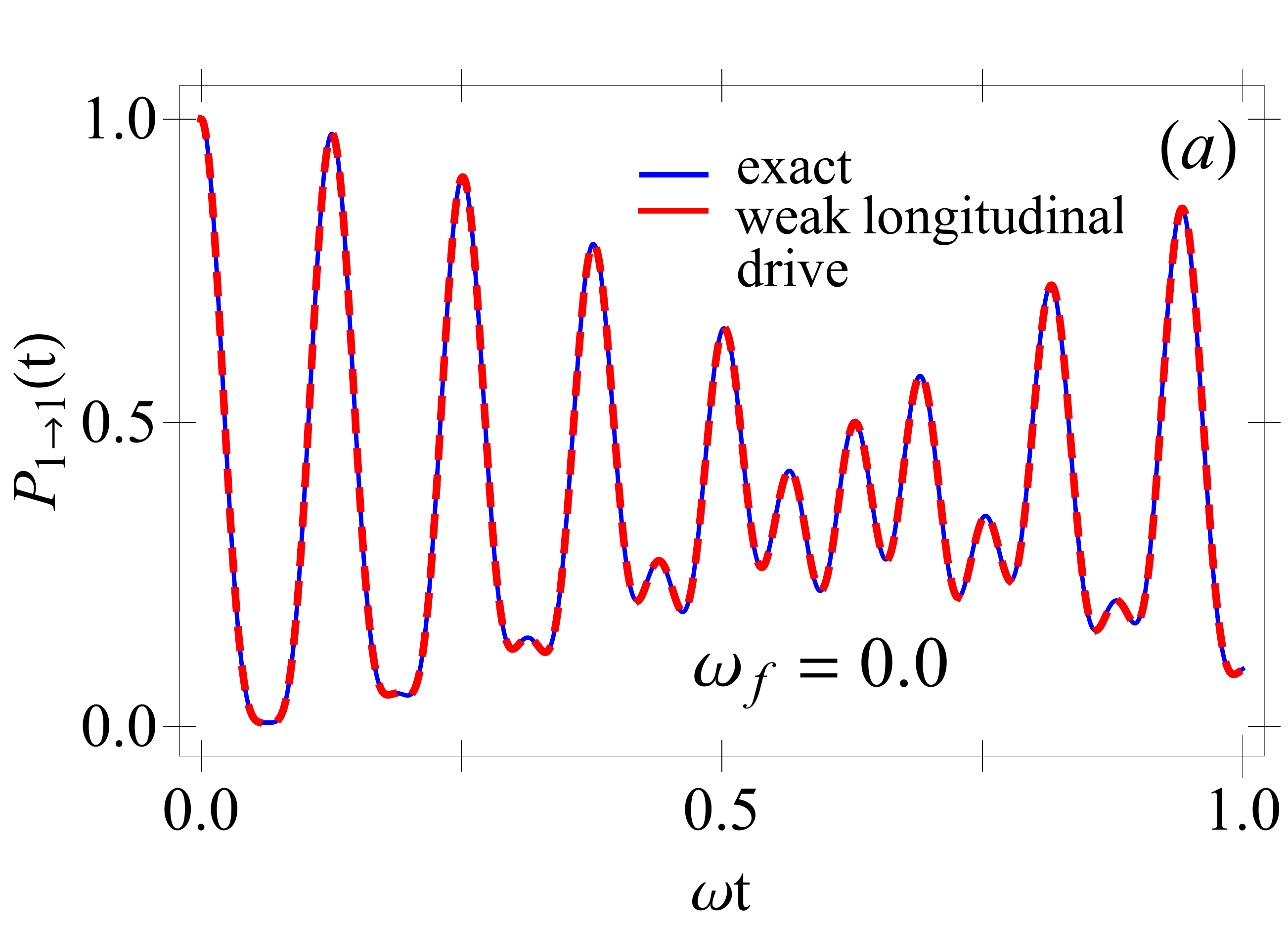}
 \includegraphics[width=4.2cm, height=4cm]{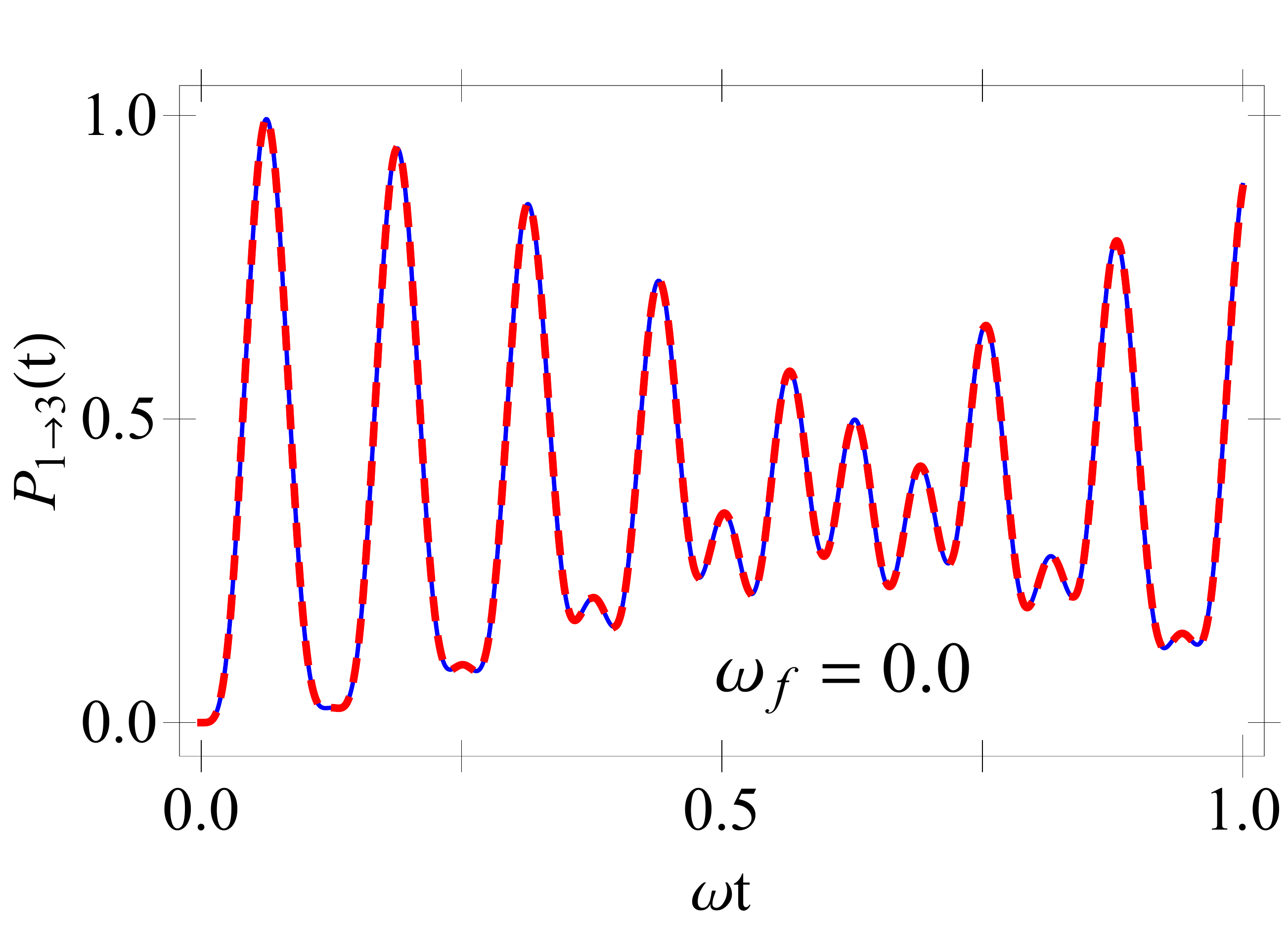}
 \end{center}
\vspace{-0.75cm}
\caption{(Color Online) Typical time-evolution of population $P_{1\to1}(t)$ returned (a) to the state $|1\rangle$ after interactions and $P_{1\to3}(t)$ transferred (b) to the state $|3\rangle$ for an initialization of the ThLS in the state with maximal spin projection $|1\rangle$. Solid blue lines are exact numerical solution to the TDSE while red dashed lines are numerical data for $P_{1\to1}(t)$ and $P_{1\to3}(t)$ respectively calculated from Eqs.(\ref{equ20a}) and (\ref{equ20d}) for $A/\omega=0.05$ (weak longitudinal driving), $\mathcal{A}_{f}/\omega=50$ and $D/\omega=5$.} \label{Figure5}
\end{figure}

\begin{figure}[!b]
\vspace{-0.5cm}
\begin{center}
\includegraphics[width=7.5cm, height=6.5cm]{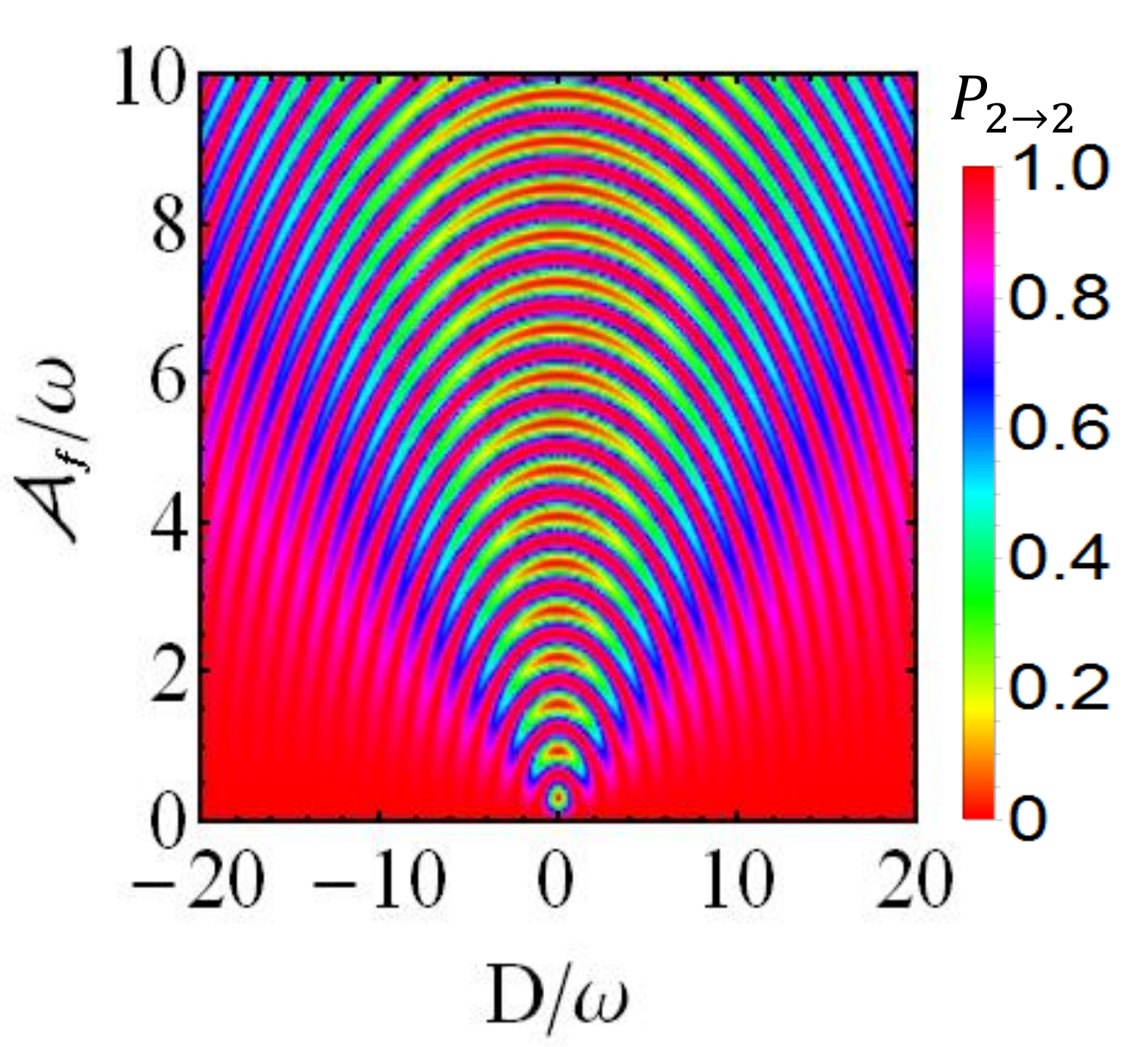}
\end{center}
\vspace{-0.75cm}
\caption{(Color Online) (Weak longitudinal driving interference patterns)  Probability $P_{2\to2}(t)$ viewed as a function of $D/\omega$ and $\mathcal{A}_{f}/\omega$ calculated from (\ref{equ20c}) for $A/\omega=0.05$, $\omega_{f}/\omega=0$ from the initial time $t_{0}=0$ to $t=5/\omega$. } \label{Figure5a}
\end{figure}

In order to obtain a simplified form  of (\ref{equ4h}) which is easier handled in this approximation, let us set $\omega_{f}=0$. Thus, the eigen-energies acquire the simplified form 
\begin{eqnarray}
\hspace{-0.5cm}E_{1}=\Omega_{0}\cot\varphi,\quad E_{2}=\Delta_{0}, \quad {\rm and} \quad E_{3}=-\Omega_{0}\tan\varphi,
\end{eqnarray}
 where 
\begin{eqnarray}
\tan2\varphi=\frac{2\Omega_{0}}{\Delta_{0}}.
\end{eqnarray}
 The eigen-vectors yield the matrix elements $w_{kj}$ . After evaluation using the familiar procedure adopted so far, this yields
\begin{eqnarray}\label{equ19cc}
\mathbf{W}=\frac{1}{\sqrt{2}}
\left[
{\begin{array}{*{20}c}
\cos\varphi &  -1 & \sin\varphi\\
-\sqrt{2}\sin\varphi & 0 & -\sqrt{2}\cos\varphi\\
\cos\varphi & 1 & \sin\varphi
\end{array} } \right].
\end{eqnarray}
The shape of this matrix suggests some relevant symmetries that will considerably reduce the length of our iterations. Thus, instead of repeating tedious calculations that finally lead to the same results, the symmetry of the first and third rows in $\mathbf{W}$ tells us that $P_{1\to1}(t)=P_{3\to3}(t)$. As an immediate consequence, $P_{1\to2}(t)=P_{2\to3}(t)$. In addition, the usual symmetries $P_{\kappa'\to\kappa}(t)=P_{\kappa\to\kappa'}(t)$ for $\kappa'\neq\kappa$ are used. So, the full matrix of transition probabilities ($3\times3$ matrix with $9$ elements) is evaluated once we know four of its matrix elements.  They are here calculated from Eq.(\ref{equ4h}) and written in explicit form as: 
\begin{subeqnarray}\label{equ20}
\nonumber P_{1\to1}(t)=\Big(\cos^{2}\Big(\frac{q_{0}t}{2}\Big)-\sin^{2}(p_{0}t)\sin^{2}\varphi\Big)^{2}\\+\frac{1}{4}\Big(\sin(q_{0}t)+\sin(2p_{0}t)\sin^{2}\varphi\Big)^{2},\slabel{equ20a}
\\\nonumber\\ 
P_{1\to2}(t)=\frac{1}{2}\sin^{2}2\varphi\sin^{2}(p_{0}t),\hspace{1.3cm} \slabel{equ20b}
\\\nonumber\\ 
 P_{2\to2}(t)=1-\sin^{2}2\varphi\sin^{2}(p_{0}t),\hspace{1cm}\slabel{equ20c}
\\\nonumber\\ 
\nonumber P_{1\to3}(t)=\Big(\sin^{2}\Big(\frac{q_{0}t}{2}\Big)-\sin^{2}(p_{0}t)\sin^{2}\varphi\Big)^{2}\\+\frac{1}{4}\Big(\sin(q_{0}t)-\sin(2p_{0}t)\sin^{2}\varphi\Big)^{2},\slabel{equ20d}
\end{subeqnarray}
where we have defined $q_{0}=\Omega_{0}\tan\varphi$ and $p_{0}=\Omega_{0}\csc2\varphi$. Making use of symmetry elucidated above, the remaining terms in the matrix of transition probabilities are deduced. These solutions are valid under the weak longitudinal driving limit and when $\omega_{f}=0$. The TDSE (\ref{equ5}) is exactly solved numerically, the results are depicted in the figure \ref{Figure5} together with our analytical results (\ref{equ20}). We see that analytical and numerical data are barely discernible confirming that (\ref{equ20}) is useful to probe ThLSs under the stated conditions. Interference patterns associated with this regime are displayed in Fig.\ref{Figure5a}.

It would be relevant to realize that at time when the fields are tuned such that $t=\pi(2N+1)/p_{0}$ ($N=0,1,2,3,...$), the matrix of transition probabilities  at points $N$ acquires the simple form
\begin{eqnarray}
\mathbf{P}_{N}=
\left[
{\begin{array}{*{20}c}
\cos^{2}\vartheta_{N} &  0 & \sin^{2}\vartheta_{N}\\
0 & 1 & 0\\
\sin^{2}\vartheta_{N} & 0 & \cos^{2}\vartheta_{N}
\end{array} } \right].
\end{eqnarray}
 Here, $\mathbf{P}_{N}\equiv\mathbf{P}(\pi(2N+1)/p_{0})$ and $\vartheta_{N}=\pi(2N+1)\sin^{2}\varphi$. For an initialization of the system in the state $|1\rangle$, paths interfere constructively when $\cos\vartheta_{N}=0$. This is achieved at $N=0$ when $\varphi=\pi/4$. 

\subsection{Strong longitudinal driving limit $A\gg\omega$}\label{SLD}

In this regime, the amplitude of the applied longitudinal $ac$ field greatly exceeds the driving field photon energy $\omega$ and the energy separation between adiabatic states such that the condition $A\omega\gg\mathcal{A}_{f}^{2}$ is always satisfied.  We observe sequential LZSM transitions when $A$ is large as compared to the uniaxis anisotropy.  The phases on the right hand side of Eq.(\ref{equ10}) have a periodic dependence. Under the stated condition, the equations in (\ref{equ9}) can readily be integrated over $2\pi/\omega$ assuming that probability amplitudes are constant. This tells us that the phases in  Eq.(\ref{equ10}) are relevant only at points where the argument of the exponential functions are stationary (resonance condition). The first equation essentially contributes when $n=-(D+\omega_{f})/\omega$ for the first exponential phase and $n=(D-\omega_{f})/\omega$ for the second exponential while the third equation is relevant when $n=(D+\omega_{f})/\omega$ for the first exponential phase and $n=(D-\omega_{f})/\omega$ for the second. The same strategy is applied to the second equation.  This leads us to 
\begin{eqnarray}\label{equ13}
i\frac{d\phi(\tau)}{d\tau}=
\left[
{\begin{array}{*{20}c}
0 &  \Omega_{-} & 0\\
\Omega_{-} & 0 & \Omega_{+}\\
0 & \Omega_{+} & 0
\end{array} } \right]\phi(\tau),
\end{eqnarray}
where
\begin{eqnarray}\label{equ14}
\hspace{-0.6cm}\Omega_{\pm}=\frac{\mathcal{A}_{f}}{2\omega\sqrt{2}}\Big[J_{\pm(D+\omega_{f})/\omega}\Big(\frac{A}{\omega}\Big)+J_{\pm(D-\omega_{f})/\omega}\Big(\frac{A}{\omega}\Big)\Big].
\end{eqnarray}
Through this, the $SU(2)$ symmetry is restored. We are interested in the instantaneous population $P_{\kappa'\to\kappa}(\tau)=|\phi_{\kappa}(\tau)|^{2}$ ($\kappa=1,2,3$) at any given time $\tau$ for the case when at time $\tau_{0}=0$ the system is in the state $|\kappa'\rangle$. Due to symmetry between levels, it is instructive to start with the case,
\begin{eqnarray}\label{equ15a}
\phi_{1}(\tau_{0})=1, \quad \phi_{2}(\tau_{0})=0, \quad \phi_{3}(\tau_{0})=0,
\end{eqnarray}
 given that other preparations are deduced by symmetry reasons. Thus, it might be interesting to realize that the Hamiltonian in Eq.(\ref{equ13}) is constant in time and that the theory constructed in subsection \ref{adiab} applies. On the other hand, the leading three-level problem is non-trivially but intimately associated with the two-level one 
$
i\dot{\mathbf{b}}(\tau)=\frac{1}{2}(
\Omega_{+}\boldsymbol{\mathrm{\sigma}}_{z}+\Omega_{-}\boldsymbol{\mathrm{\sigma}}_{x})\mathbf{b}(\tau),
$
such that 
\begin{subeqnarray}\label{equ15b}
&\phi_{1}(\tau)=|b_{1}(\tau)|^{2}-|b_{2}(\tau)|^{2},\slabel{equ15ca}
\\\nonumber\\
&\phi_{2}(\tau)=b_{1}^{*}(\tau)b_{2}(\tau)-b_{1}(\tau)b_{2}^{*}(\tau),\slabel{equ15cb}
\\\nonumber\\
&\phi_{3}(\tau)=b_{1}^{*}(\tau)b_{2}(\tau)+b_{1}(\tau)b_{2}^{*}(\tau),\slabel{equ15cc}
\end{subeqnarray}
where $\mathbf{b}(\tau)=[b_{1}(\tau), b_{2}(\tau)]^{T}$ is a two-component vector transition amplitude. The Hamiltonian in this auxiliary problem describes a two-level atom undergoing Rabi oscillations\cite{Ansari}. The difference between the frequency of the external fields and the Bohr transition frequency of the system is constant as well as interactions between levels. After solving the intermediate problem, and considering the connecting relation in Eqs.(\ref{equ15ca})-(\ref{equ15cc}), one finds that:
\begin{subeqnarray}\label{equ15c}
& \phi_{1}(\tau)=\dfrac{\Omega_{+}^{2}}{\Omega^{2}_{\rm eff}}+\dfrac{\Omega_{-}^{2}}{\Omega^{2}_{\rm eff}}\cos[\Omega_{\rm eff}\tau],
\\\nonumber\\
&\phi_{2}(\tau)=-\dfrac{i\Omega_{-}}{\Omega_{\rm eff}}\sin[\Omega_{\rm eff}\tau],
\\\nonumber\\
&\phi_{3}(\tau)=\dfrac{\Omega_{+}\Omega_{-}}{\Omega^{2}_{\rm eff}}-\dfrac{\Omega_{+}\Omega_{-}}{\Omega^{2}_{\rm eff}}\cos[\Omega_{\rm eff}\tau],
\end{subeqnarray}
where $\Omega_{\rm eff}=[\Omega_{+}^{2}+\Omega_{-}^{2}]^{1/2}$ is the total Rabi frequency. We can evaluate the populations $P_{1\to1}(\tau)=|\phi_{1}(\tau)|^{2}$,  $P_{1\to2}(\tau)=|\phi_{2}(\tau)|^{2}$ and $P_{1\to3}(\tau)=|\phi_{3}(\tau)|^{2}$.  They are periodic $P_{\kappa'\to\kappa}(\tau+2\pi/\Omega_{\rm eff})=P_{\kappa'\to\kappa}(\tau)$. It can be verified that the system remains conservative throughout the course of time. Therefore, by adding together the three probabilities calculated from Eq.(\ref{equ15c}) one  gets $1$.  

\begin{figure}[!h]
\vspace{-0.5cm}
\begin{center}
 \includegraphics[width=7.2cm, height=5cm]{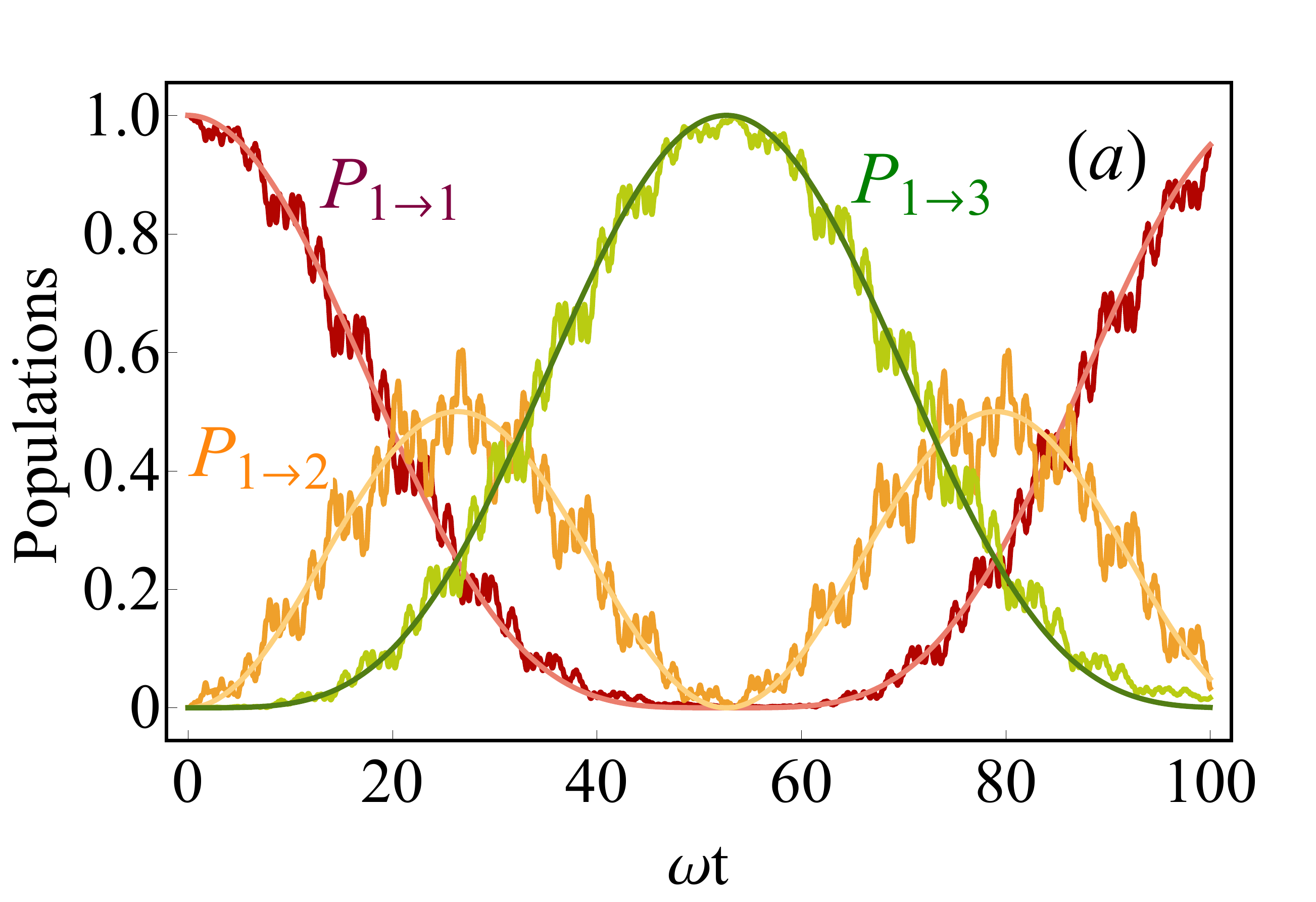}\\\vspace{-0.25cm}
 \includegraphics[width=7.2cm, height=5cm]{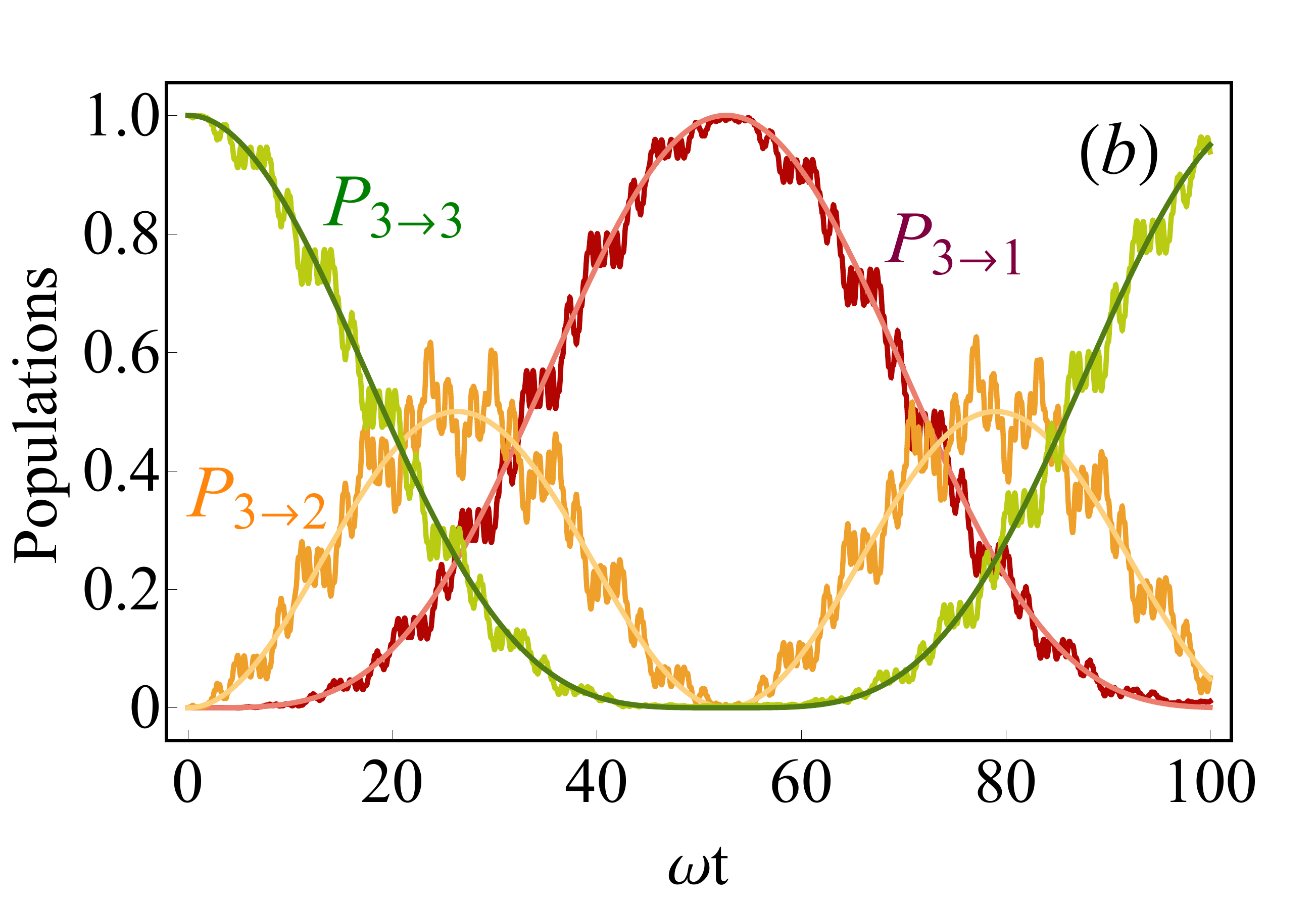}
\end{center}
\vspace{-0.65cm}
\caption{(Color Online) Typical time-evolution of population on diabatic levels calculated in the strong longitudinal driving limit ($A\gg\omega$).  LZSM oscillations are exact results obtained by numerically solving the TDSE. Solid lines without LZSM oscillations are analytical data calculated from Eq.(\ref{equ15d}). To plot all graphs, we have used $A/\omega=10$, $\mathcal{A}_{f}/\omega=0.5$, $D/\omega=5$ and $\omega_{f}=10\omega$. In this regime as indicated earlier, we have cascaded LZSM oscillations in the population of level. The time is in the unit of $1/\omega$. } \label{Figure4}
\end{figure}

\begin{figure}[]
\vspace{-0.75cm}
\begin{center}
 \includegraphics[width=7cm, height=5cm]{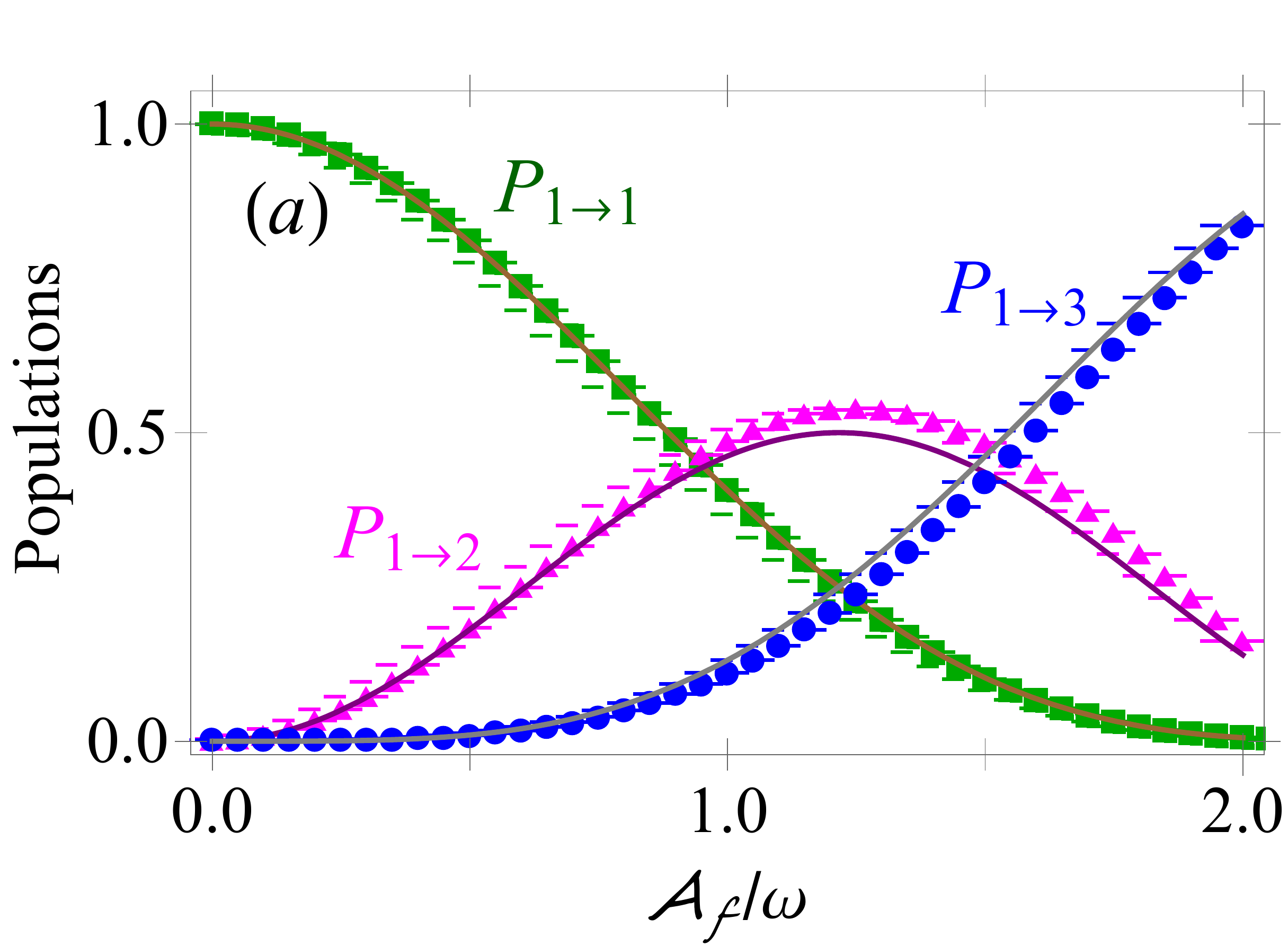}\\\vspace{-0.25cm}
 \includegraphics[width=7cm, height=5cm]{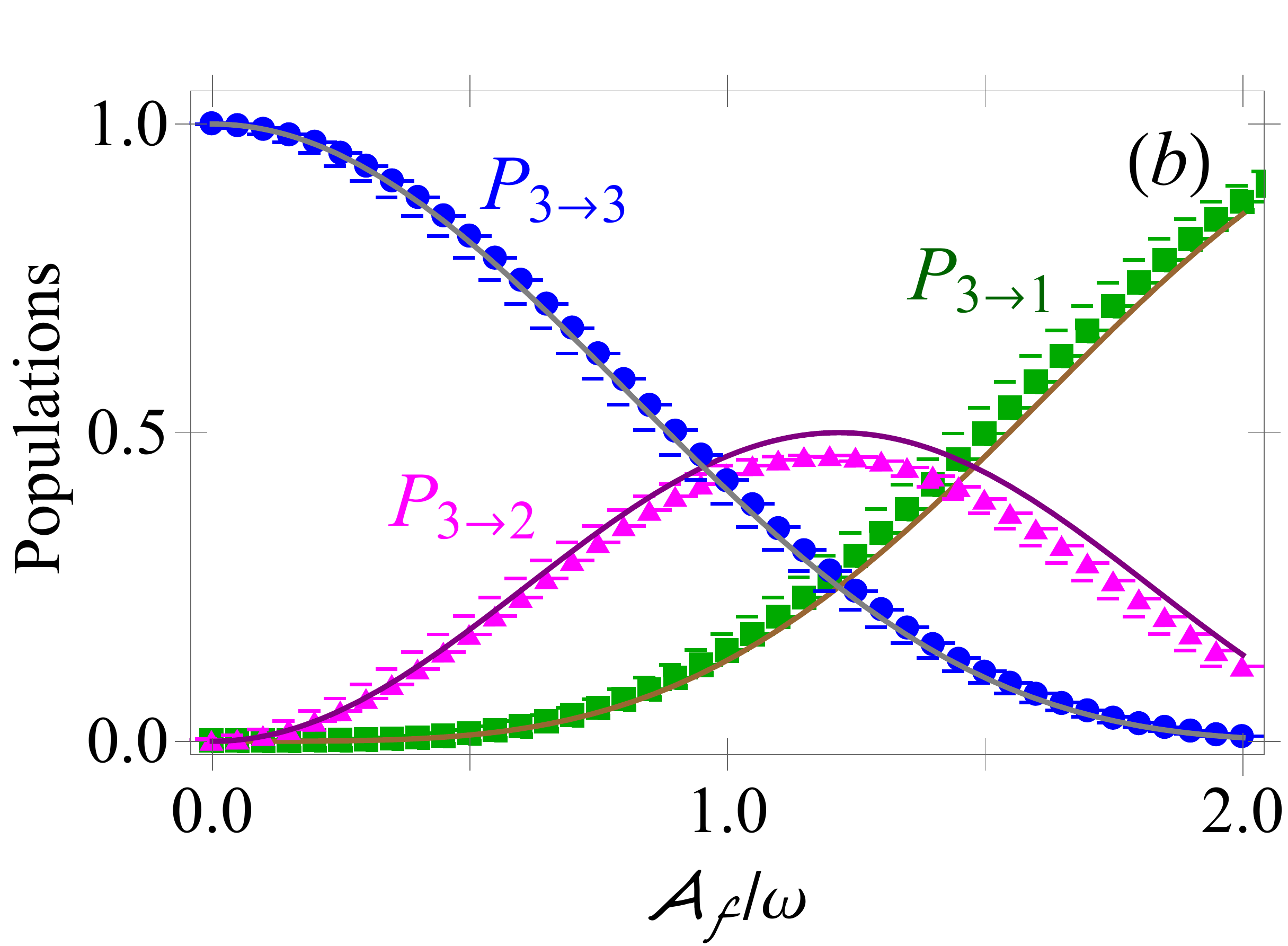}
\end{center}
\vspace{-0.75cm}
\caption{(Color Online) Numerical solutions of Eq.(\ref{equ3}) with the model (\ref{equ4}) and probabilities $P_{\kappa'\to\kappa}(\tau)$ calculated from Eq.(\ref{equ15d}) at time $\tau=10$ with $A/\omega=30$ and $D/\omega=3$. Solid lines are analytical results while symbolic lines are numerical results. These figures indicate  a suppression of tunneling at $\mathcal{A}_{f}/\omega=0$. Thus, the CDT is defined by the longitudinal drive and suppressed by the transverse drive in this case.} \label{Figure8}
\end{figure}

We can now consider other initial preparations of the ThLS: the cases when the system is initialized in the diabatic states $|2\rangle$ or $|3\rangle$. The strategy used for the previous case applies as well. However, we exploit the symmetry between levels and construct the following transition matrix: 
\begin{equation} \label{equ15d} 
\mathbf{P}_{\kappa\to \kappa'} (\tau)=\left[\begin{array}{ccc} {|\phi_{1} (\tau)|^{2} } & {|\phi_{2} (\tau)|^{2} } & {|\phi_{3} (\tau)|^{2} } \\ {|\phi_{2} (\tau)|^{2} } & {1-2|\phi_{2} (\tau)|^{2} } & {|\phi_{2} (\tau)|^{2} } \\ {|\phi_{3} (\tau)|^{2} } & {|\phi_{2} (\tau)|^{2} } & {|\phi_{1} (\tau)|^{2} } \end{array}\right]. 
\end{equation} 
 The element in position ($\kappa',\kappa$) represents the probability $P_{\kappa'\to \kappa}(\tau)$. We have equally compared the results of this subsection with numerics (see figures \ref{Figure4} and  \ref{Figure8}). 

When the time scale of the process is small as compared to oscillation time $\tau\ll\Omega^{-1}_{\rm eff}$, the probability amplitudes are no longer functions of the total Rabi frequency and read $\phi_{1}(\tau)\approx 1-(\Omega_{-}\tau)^{2}/2$, $\phi_{2}(\tau)\approx -i\Omega_{-}\tau$ and $\phi_{3}(\tau)\approx (\Omega_{+}\Omega_{-}\tau)^{2}/2$. The system mainly remains in its initial state. Thus, the action of the longitudinal drive in the strong driving limit becomes noticeable after the characteristic time $\tau_{\omega}=\Omega^{-1}_{\rm eff}$. This is verified by numerically solving the TDSE in the strong longitudinal driving limit. The exact solutions are displayed in the figures \ref{Figure4} and \ref{Figure8} together with analytical results (\ref{equ15d}) for an initialization of the system in the diabatic states $|1\rangle$ and $|3\rangle$. A satisfactory agreement is to be noted. Figure \ref{Figure8} however brings out crucial information. Indeed, on the figure therein, $0\le\mathcal{A}_{f}/\omega\le2$, it appears that as soon as $\mathcal{A}_{f}/\omega$ exceeds $1$, there is a slight but noticeable deviation between analytical and numerical results especially for $P_{1\to2}(t)$ and $P_{1\to3}(t)$. This is clearly an indication that in addition to $A/\omega\gg1$ one must compulsorily fix $\mathcal{A}_{f}/\omega<1$ to find a concordance between the results of this subsection  and exact results. Then, our analytical results globally hold when two conditions are satisfied: $A/\omega\gg1$ and $\mathcal{A}_{f}/\omega<1$ for arbitrary $\tau$, $\omega_{f}/\omega$ and $D/\omega$.  

Thus, in the limit of exceedingly large longitudinal driving amplitudes $A/\omega\gg1$, such frequencies exhibit periodic dependence and are comparable to
\begin{eqnarray}\label{equ14a}
\Omega_{\pm}\approx\sqrt{\frac{\mathcal{A}_{f}^{2}}{\pi A\omega}}\cos\Big(\frac{A}{\omega}\mp\frac{\pi D}{2\omega}-\frac{\pi}{4}\Big)\cos\Big(\frac{\pi\omega_{f}}{2\omega}\Big).
\end{eqnarray}
Thus, when the fields are tuned such that $\omega_{f}=(2N+1)\omega$ ($N=0,1,2,3,...$), Rabi frequencies all cancel out. Diabatic states cannot communicate and there is a population trapping. The actions of the fields are mutually inhibited. 

It is worth nothing that one can switch off the transverse drive frequency ($\omega_{f}=0$) and maintain its amplitude $\mathcal{A}_{f}$ constant such that the inter-level distance between level positions  remains constant in time and never turns off. This situation is encountered in versatile  experiments and Rabi frequencies (\ref{equ14}) are $\Omega_{\pm}=(\mathcal{A}_{f}/\omega\sqrt{2})J_{\pm D/\omega}(A/\omega)$.

\section{More approximations}\label{Sec3}

\subsection{On-resonance fields $\omega_{f}=\omega$}

\begin{figure}[!h]
\begin{center}
\includegraphics[width=7.5cm, height=5.7cm]{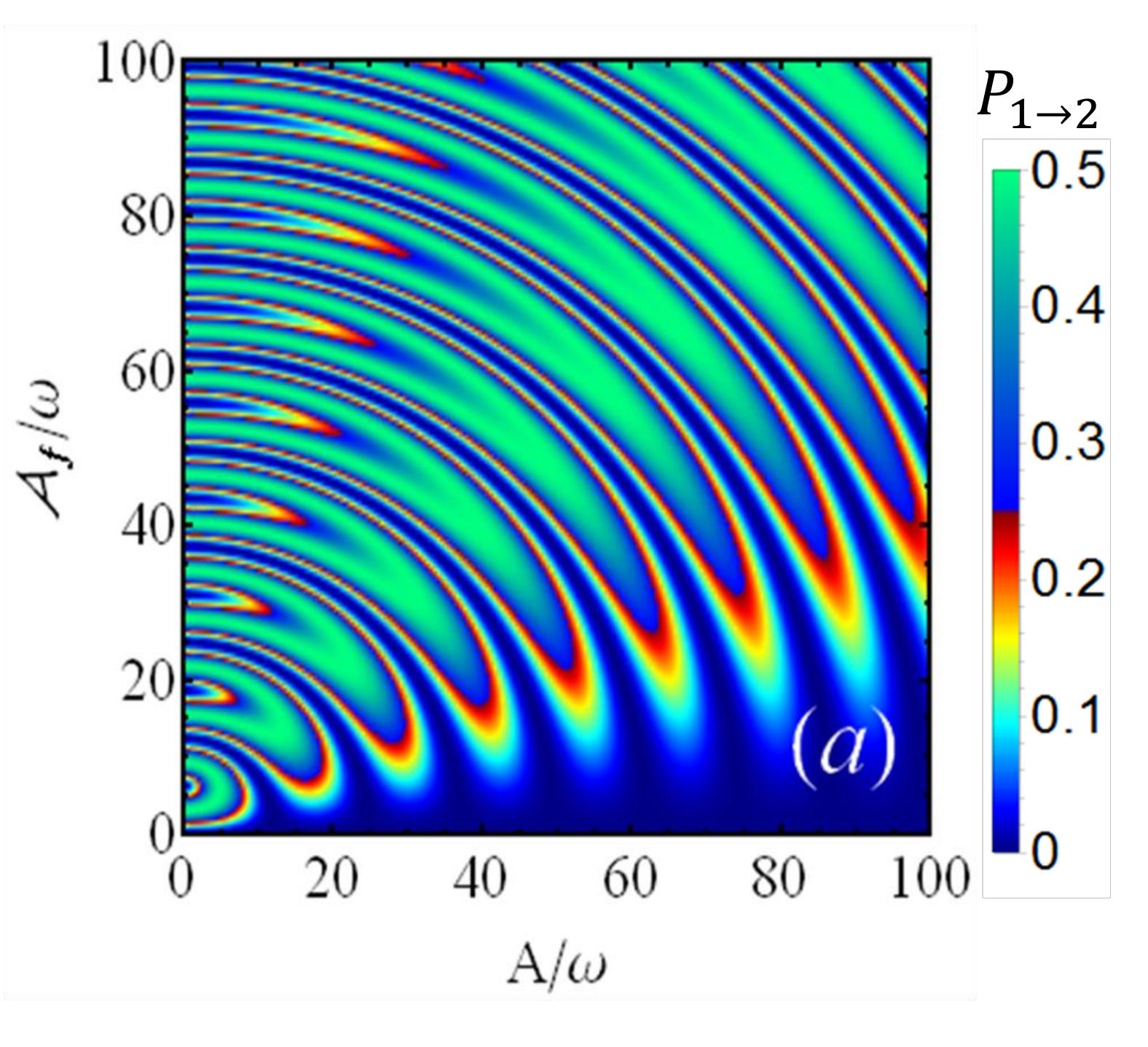}\vspace{-0.3cm}
\includegraphics[width=7.5cm, height=5.7cm]{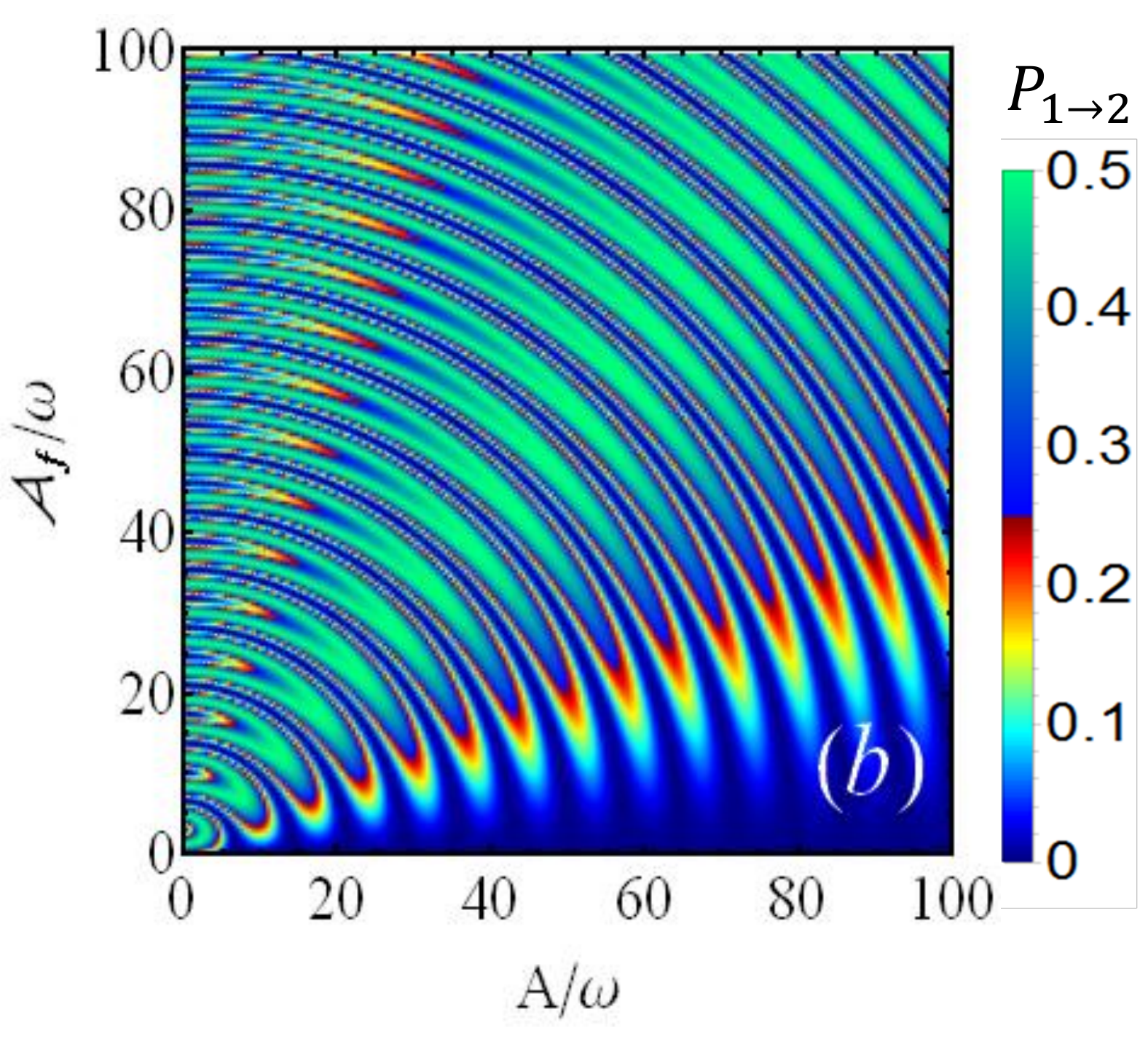}
\end{center}
\vspace{-0.8cm}
\caption{ (Color Online) (a): Excited-state probability $P_{1\to2}$ versus the amplitude $A/\omega$ and $\mathcal{A}_{f}/\omega$ for $t=100/\omega$; (b): same as (a) for $t=200/\omega$. We observed that as the periodic drive lasts long, the ThLS splits and recombines several times and the number of fringes increase.}
\label{Figure10}
\end{figure}

 It might also be interesting for technical purposes to consider the case when the longitudinal and transverse $ac$ drives are in resonance with each other i.e. synchronized such that $\omega_{f}=\omega$ and there is no static shift in the detuning ($D=0$). This leads to an interesting dynamic with $SU(2)$ symmetry which is exactly elucidated without resorting any approximation. The generic periodic (with periodicity $2\pi/\omega$) Hamiltonian describing this situation reads
\begin{eqnarray}\label{equation1} 
\mathcal{H}_{SU(2)}(t)=A\cos (\omega t)S^{z}+\mathcal{A}_{f}\cos (\omega t)S^{x}.
\end{eqnarray}
 For a dynamical description of the model, an equation similar to (\ref{equ5}) is written down and interpreted as the kernel of the problem. It is solved by adopting the change of variable  $z=\sin\omega t$ which considerably simplifies the  task leading us to the Rabi problem, 
\begin{eqnarray}\label{equation2} 
\mathcal{H}_{SU(2)}=\frac{A}{\omega}S^{z}+\frac{\mathcal{A}_{f}}{\omega}S^{x}.
\end{eqnarray} 
 The method presented in subsection \ref{adiab} is valid insofar as $\mathcal{H}_{SU(2)}$ is constant in time. The eigen-energies of this last Hamiltonian are $E_{1,3}=\pm(\mathcal{A}_{f}/\omega)\csc(2\vartheta)$ and $E_{2}=0$ where the index $1,2$ and $3$ are ascribed to the states $|1\rangle$, $|2\rangle$ and $|3\rangle$ with $\tan2\vartheta=-\mathcal{A}_{f}/A$. The eigen-vectors are calculated following the  familiar procedure adopted so far. This allows us to construct the rotation matrix $\mathbf{W}=e^{2i\vartheta S_{y}}$. Exact solutions to this problem are of the form (\ref{equ4h}) where $w_{11}=w_{33}=\cos^{2}\vartheta$, $w_{13}=w_{31}=\sin^{2}\vartheta$, $w_{12}=w_{23}=\sin2\vartheta/\sqrt{2}$ and $w_{21}=w_{32}=-\sin2\vartheta/\sqrt{2}$. Thus,
\begin{eqnarray}\label{equ31}
\nonumber P_{\kappa'\to \kappa}(t)=\Big[w_{\kappa'1}w_{\kappa1}\hspace{4cm}\\\nonumber+w_{\kappa'2}w_{\kappa2}\cos[(\mathcal{A}_{f}/\omega)\csc(2\vartheta)\sin(\omega t)]\\\nonumber+w_{\kappa'3}w_{\kappa3}\cos[(2\mathcal{A}_{f}/\omega)\csc(2\vartheta)\sin(\omega t)]\Big]^{2}\\\nonumber+
\Big[w_{\kappa'2}w_{\kappa2}\sin[(\mathcal{A}_{f}/\omega)\csc(2\vartheta)\sin(\omega t)]\\+w_{\kappa'3}w_{\kappa3}\sin[(2\mathcal{A}_{f}/\omega)\csc(2\vartheta)\sin(\omega t)]\Big]^{2}.
\end{eqnarray}
This solution is exact and holds for arbitrary parameters $A$, $\mathcal{A}_{f}$, and $\omega$. Let us now consider the special case when the readout of the ThLS is performed at time $t=\pi N/\omega$ (where $N=0,1,2,3,...$). The results of the measurement reveal that the system is completely returned to its original diabatic state $|\kappa'\rangle$. In this situation, Eq.(\ref{equ31}) acquires the form
\begin{eqnarray}\label{equ32}
P_{\kappa'\to \kappa}\Big(\frac{\pi N}{\omega}\Big)=\Big[\sum_{\ell=1}^{3}w_{\kappa'\ell}w_{\kappa\ell}\Big]^{2}.
\end{eqnarray}
Using the properties of $w_{nk}$ given in Eq.(\ref{aa6}), one can verify that in this situation, the occupation probability $P_{\kappa'\to \kappa'}(\pi N/\omega)=1$ while the transition probabilities $P_{\kappa'\to \kappa}(\pi N/\omega)=0$ (for $\kappa\neq\kappa'$), and there is a complete population return. Interactions between paths  are destructive. This is not only the consequence of the $SU(2)$ symmetry between adiabatic states with extremal spin projections whose energies are antisymmetric $E_{1}=-E_{3}$ and the fact that the middle state is a dark state (state with zero energy) but also the major fact that at instants $\pi N/\omega$, the longitudinal and transverse drives are on-resonance with the same frequency. Such a result is not trivial when the   $SU(3)$ symmetry is preserved. The interference patterns associated with this situation are presented in  Fig.\ref{Figure10}. This figure indicates an increase of the number of fringes as the driving process lasts longer.  The upper panel corresponds to a process stopped as $t=100/\omega$ while the lower panel corresponds the one terminated at $t=200/\omega$. The number of fringes significantly increases.  

\subsection{Adiabatic $SU(3)$ LZSM interferometry}

\begin{figure}[!h]
\vspace{-0.5cm}
\begin{center}
 \includegraphics[width=4cm, height=4cm]{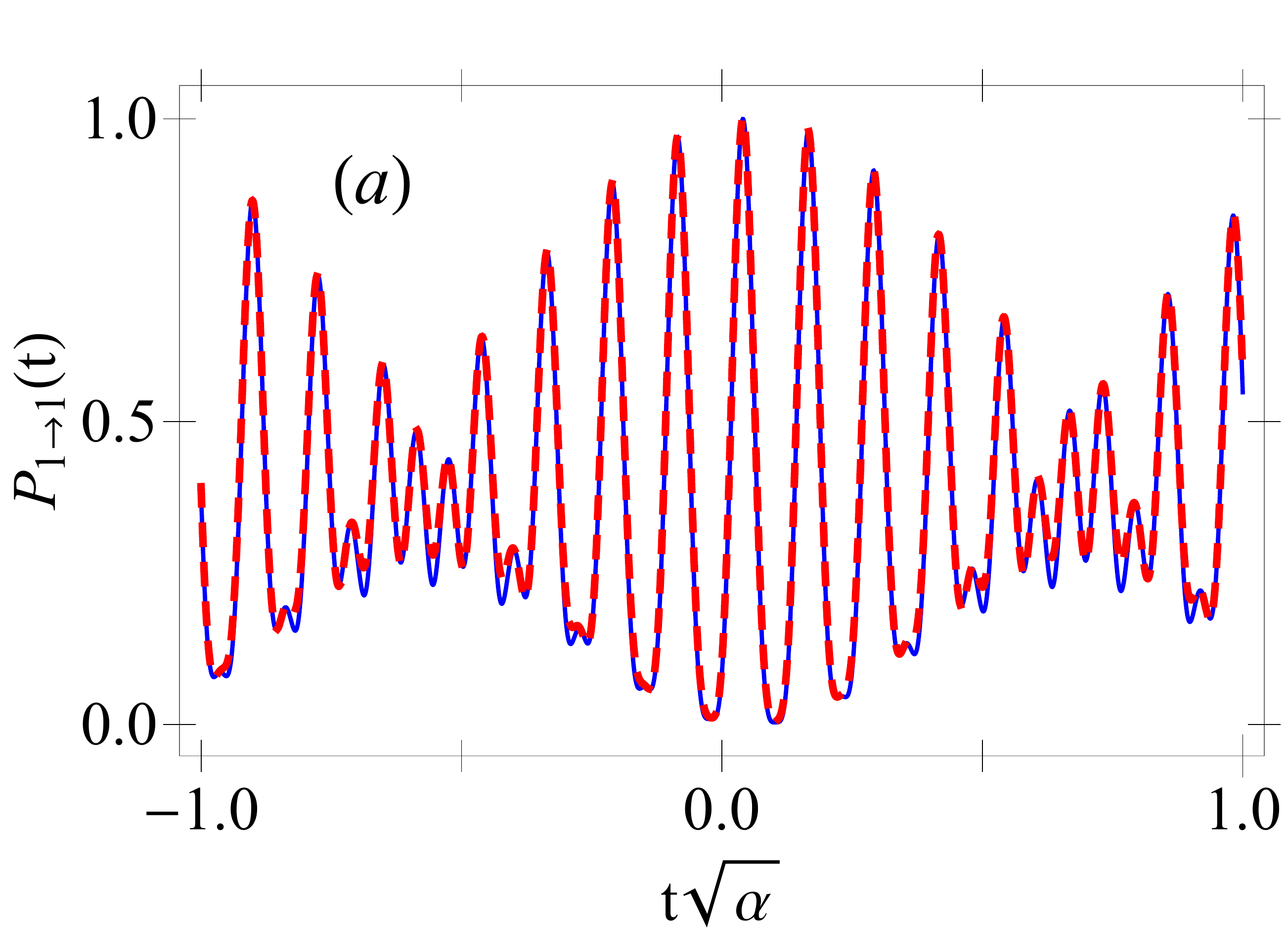}
 \includegraphics[width=4cm, height=4cm]{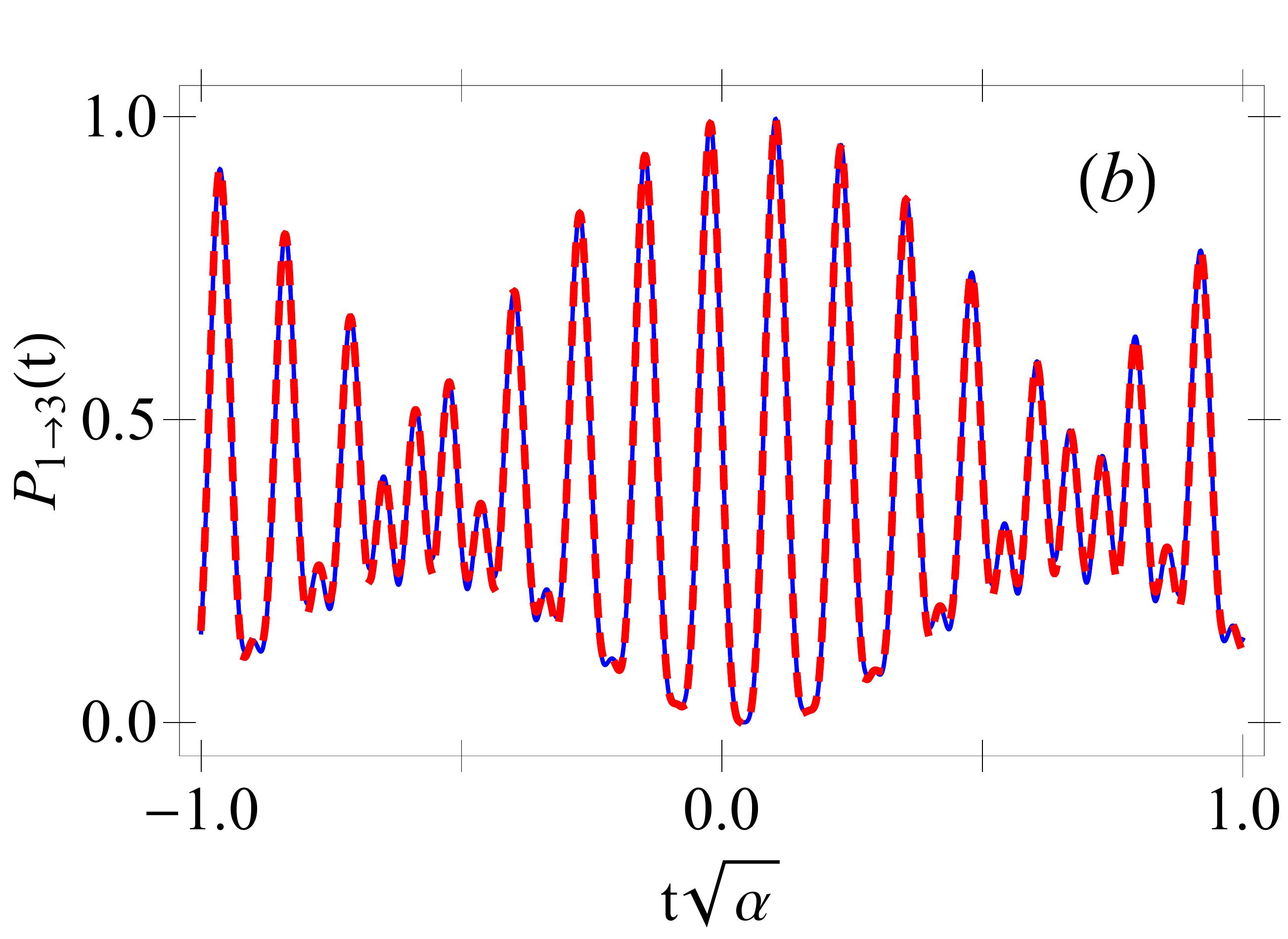}\\\vspace{-0.25cm}
\includegraphics[width=4cm, height=4cm]{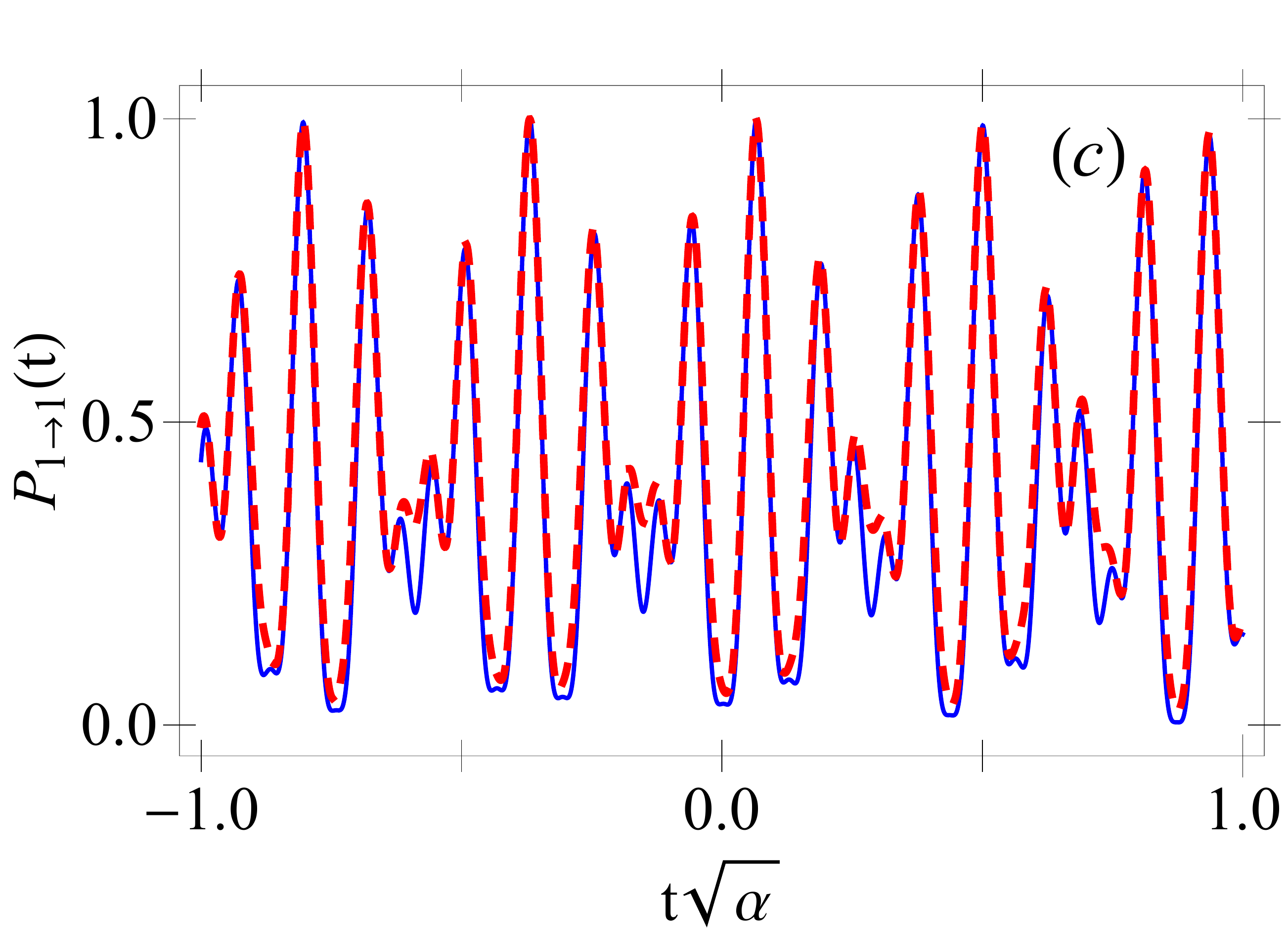}
 \includegraphics[width=4cm, height=4cm]{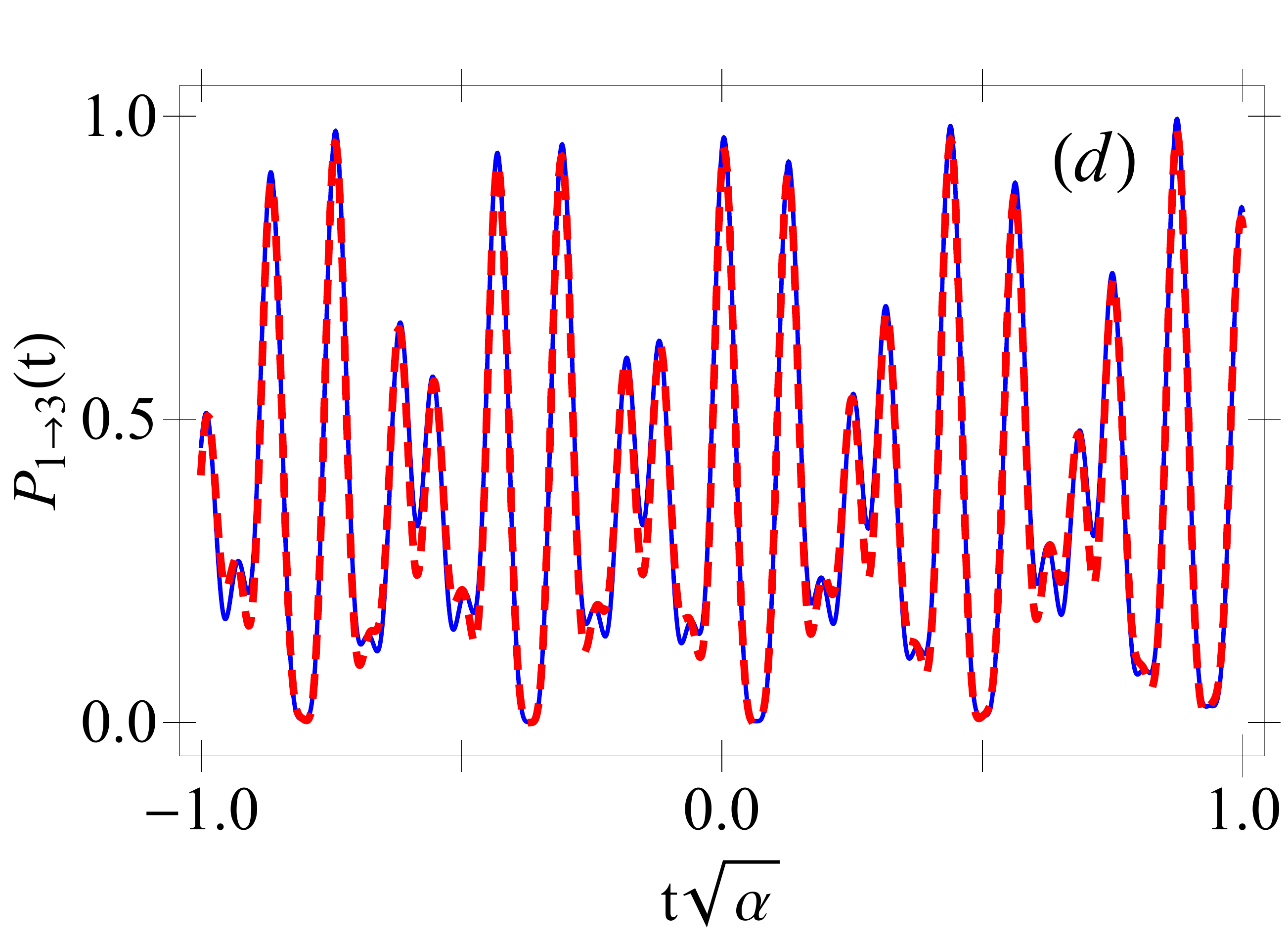}
 \end{center}
\vspace{-0.75cm}
\caption{ (Color Online) Numerical solutions of Eq.(\ref{equ3}) with the model  $\mathcal{H}_{SU(3)}(t)=\alpha t S^{z}+\Delta S^{x}+D(S^{z})^{2}$ and transition probabilities $P_{1\to1}(t)$ and $P_{1\to3}(t)$ calculated from the general formula Eq.(\ref{equ4h}). To calculate all graphs we have considered the initial time $t_{0}=-10$ (as our solution Eq.(\ref{equ4h}) works for arbitrary initial time), the coupling $\Delta=50$, the sweep velocity $\alpha=0.05$. Solid (blue) lines are exact numerical results while dashed (red) lines are analytical results. On the upper panel (a) and (b), we have taken $D=5$ while for the lower panel (c) and (d), $D=15$. This is another evidence that our results are valid for arbitrary $D$.} \label{Figure11}
\end{figure}

To demonstrate once again the efficiency of our adiabatic treatment, let us consider in conclusion the non-resonant three-level $SU(3)$ LZSM model introduced in Ref.\onlinecite{Ken2013}. Here, the detuning linearly depends on the time and the Rabi frequency is constant and never turns off. This model stems from (\ref{equ2}) by switching off the frequency of the transverse drive ($\omega_{f}=0$) and linearizing the longitudinal drive at the vicinity of the point $t=t'+\arccos(-D/A)/\omega$ (where the states with $m=+1$ and $m=0$ come close, see Figure \ref{Figure0}) such that $\omega t'\ll1$. After performing these actions and shifting the time as $t''=t'-D/\alpha$, we keep the notation $t$ for the time instead of $t''$ and arrive at 
\begin{eqnarray}\label{equation3} 
\mathcal{H}_{SU(3)}(t)=\alpha t S^{z}+\Delta S^{x}+D(S^{z})^{2},
\end{eqnarray} 
where $\alpha=A\omega\sqrt{1-(D/A)^{2}}$ is the sweep rate of the control protocol and $\Delta=\mathcal{A}_{f}$, is the tunnel amplitude. In Ref.\onlinecite{Ken2013}, the model is discussed  in the non-adiabatic limit $\Delta^{2}\ll\alpha$. Here, with the help of the theory elaborated  in section \ref{adiab}, we claim that the formula Eq.(\ref{equ4h}) perfectly works for the $SU(3)$ LZSM model when $\Delta^{2}\gg\alpha$ (condition for adiabatic evolutions) with $\omega_{\pm}(t)=\pm\alpha t+D$, $\Delta_{ij}(t)=\Delta/\sqrt{2}$ and $\Delta_{ii}(t)=0$ $(i,j)=1,2,3$. Evidence of our assertions is depicted on figure \ref{Figure11} where numerical and analytical results are simultaneously displayed. We merely see that both the graph for analytical and numerical calculations are barely discernible. Thus, when the condition $\Delta^{2}\gg\alpha$ is realized, adiabatic states $|\varphi_{\kappa}(t)\rangle$ slowly transport populations from one diabatic state to another at avoided level crossings. There is no direct transitions between adiabatic states and consequently no mixture of populations at avoided level crossing. So, non-adiabatic paths do not interfere. What are  finite-time populations on diabatic levels? The formula (\ref{equ4h}) answers this question for all initial preparation of the system.

\section{Applications to Quantum information processing}\label{Sec4}

Recent breakthroughs in QIP have classified the NVC in diamond as a good platform for implementing logical gates and developing quantum technologies\cite{Zhou, Du,  comment1, Ansari, Fuchs, Wubs}. This is most likely because of its spin-$1$ ground-state\cite{Zhou, Du,  comment1, Ansari, Fuchs, Wubs} which can be located using confocal microscopy and manipulated using gates voltages\cite{Fuchs}. In addition, its spin is easily initialized to the ground-state at room temperature by optical pumping and read out through spin dependent photoluminescence measurement\cite{Fuchs}. As another great advantage which makes this yet an attractive system for both fundamental investigations of quantum behavior and a good candidate for QIP, the spin of the NVC is hosted in a nearly spinless lattice of diamonds and therefore possesses a  long coherence time at room temperature.  The corresponding dynamics may be described by the model in Eq.(\ref{equ1}) when $D=2\pi\times2.88$GHz neglecting hyperfine interactions between spin states and the Nitrogen\cite{Zhou, Du,  comment1, Ansari, Fuchs, Wubs}. The quantization axis is the axis between the Nitrogen and the Vacancy. Remarkably, for optimal control\cite{Dolde}, the Hamiltonian of a driven nucleus  in the diamond can be written as $\mathcal{H}(t)=g_{e}\mu_{B}\mathbf{B}(t)\cdot\mathbf{S}+D(S^{z})^{2}$ where $g_{e}$ is the electron gyromagnetic factor, $\mu_{B}$ the Bohr magneton. $\mathbf{B}(t)=A\cos(\omega t)e_{z}+\mathcal{A}_{f}\cos(\omega_{f} t)e_{x}$ is the control magnetic field, $\mathbf{S}=S^{x}e_{x}+S^{y}e_{y}+S^{z}e_{z}$ and $e_{x,y,z}$ are unit vectors representing the polarization of the control signal. This approximation is well justified given that most experiments are performed at low magnetic fields, and then the dominant interaction is provided by the zero-field splitting  term $D$. Our results immediately apply by rescaling $A\to g_{e}\mu_{B} A$ and $\mathcal{A}_{f}\to g_{e}\mu_{B} \mathcal{A}_{f}$ everywhere they appear in our equations. Importantly, there is no need to apply an additional static magnetic field to lift the degeneracy between the states with $m=-1$ and $m=+1$ of the NV in order to reduce its dynamics to that of a TLS, as it was done in Refs.\onlinecite{Zhou, Du,  comment1, Ansari, Fuchs, Wubs, Danon, Barfuss, child}. Our results merely apply by simply respecting the conditions of validity specified for each of the cases discussed. 

\begin{figure}[]
\vspace{-0.25cm}
\begin{center}
\includegraphics[width=7.5cm, height=6.5cm]{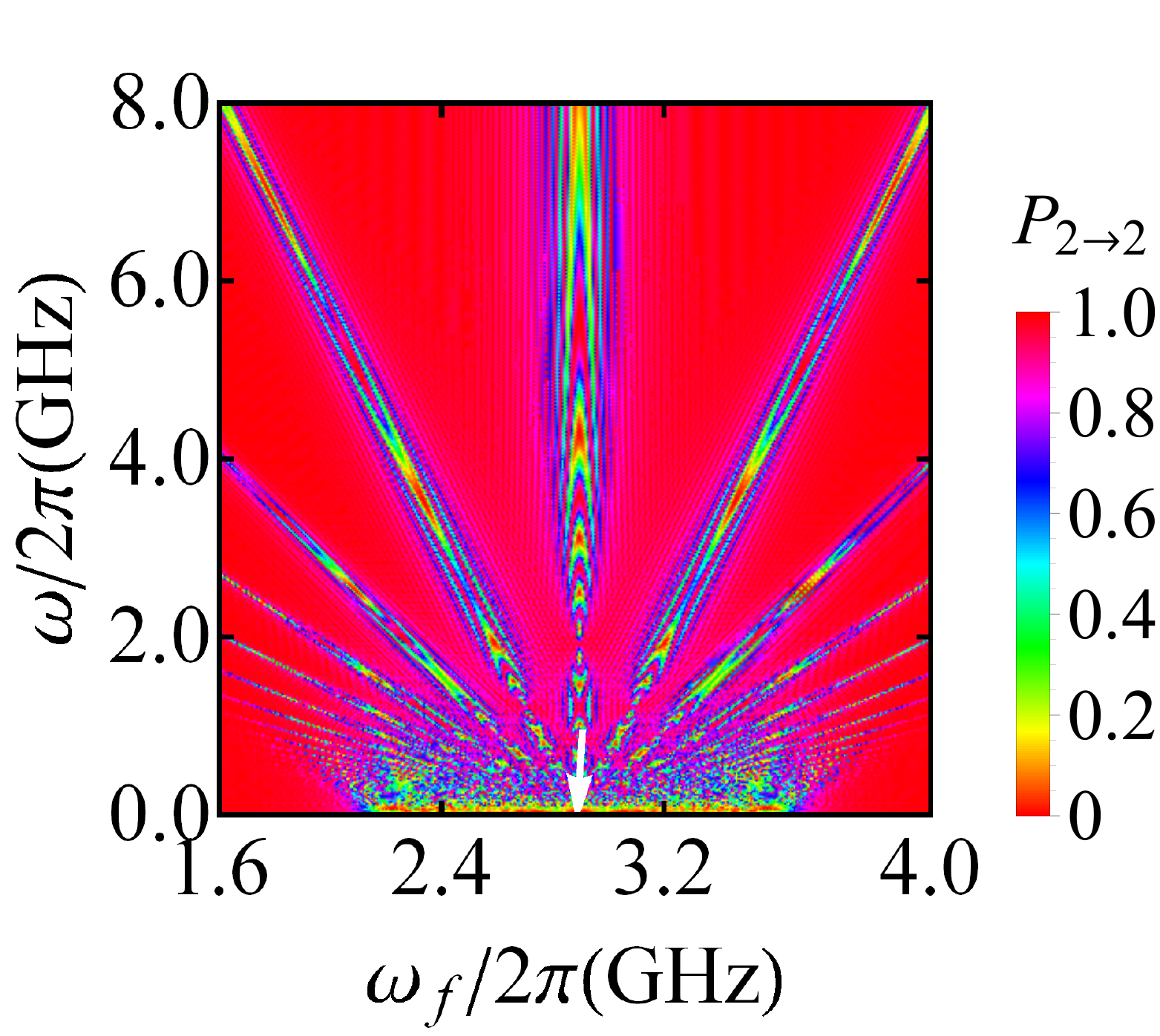} 
\end{center}
\vspace{-0.75cm}
\caption{(Color Online) Population  $P_{2\to2}(t)=|C_{2}(t)|^{2}$ remained on the diabatic state $|2\rangle$ at time $t=7.96$ns for an initial preparation of the NVC spin ground-state in the state $|2\rangle$ at $t_{0}=0.0$. It is calculated by numerically solving the TDSE (\ref{equ5}) with the model (\ref{equ1}). For numerical implementation,  we have considered $A/2\pi=4.6$GHz and $\mathcal{A}_{f}/2\pi=0.6$GHz and  $D=(2\pi)2.88$GHz. The vertical arrow points on $D$. Note that Eq.(\ref{equ11a}) reproduces a similar result.}\label{figure12}
\end{figure}

In order to prove the efficiency of our theoretical treatment, let us compare our results with those of the experiment in Ref.\onlinecite{child}  conducted at room temperature with a combination of a Micro-Wave (MW) and a Radio-Frequency (RF) in an NVC in diamond and explained numerically and analytically with the aid of a two-state model Hamiltonian. To this end, the longitudinal and transverse drives respectively match the MW and RF signals used in that experiment. They are oriented such that with the quantization axis of the NVC, they form a two-dimensional orthogonal system. This allows the control and manipulation of the NVC spin ground-state. Remark, the experiment is conducted  in the regime of weak MW and strong RF and may thus be associated with the weak transverse drive regime discussed in Subsection \ref{diab}. In Ref.\onlinecite{child},  a two-state model is used to explain the experimental results. However, no analytic formula (for populations) which mimics experimental results is presented. We have found that  $P_{2\to2}(t)\approx 1-p_{+}(t)-p_{-}(t)$ in Eq.(\ref{equ11a}) reproduce the results of the experiment quite well, given that the data calculated from this formula are in satisfactory qualitative agreement. These observations bring out two important pieces of information. Firstly,  the model (\ref{equ1}) can describe the experiment and secondly, our  analytical and numerical treatments are well applicable in NVCs and consequently in QIP. In order to confirm the first piece information, we have plotted Fig.\ref{figure12} which is qualitatively comparable with those in Ref.\onlinecite{child}.  Therefore, without necessary reducing the dynamics of the NVC (three-level system)  to that of a two-state system, we believe that experiments in NVC and in many other doubly periodically driven three-state systems will be explained with the aid of the results in this paper. Thus, questions of optimal control of qutrit may be easily addressed both qualitatively and quantitatively for a breakthrough towards the coveted quantum computer. 

\section{Conclusions}\label{Sec5}

We have investigated a class of three-level systems (ThLS) in which both the detuning and transverse tunnel amplitudes are periodically modulated by classical fields. Our treatments are based on several approximations: the weak and strong longitudinal and transverse drive limits obtained by comparing the characteristic frequency of the longitudinal drive with the fields' amplitudes. The ranges of validity and relevance of our approximations are verified by numerical tests. We obtained satisfactory agreement between analytical and numerical data for each approximation considered. The expressions for populations on levels adhere with the gross profile of exact solutions.  The results presented in this research work offer a solid background to discuss the dynamics of a periodically driven three-state Nitrogen-Vacancy Center without necessary applying an additional magnetic field which lifts the degeneracy between the states $m=0$ and $m=\pm1$ as it was done in Refs.\onlinecite{ Du, Zhou,  comment1, Ansari, Fuchs, Wubs}. They are compared with experimental results in Ref.\onlinecite{child} with a satisfactory qualitative agreement. The listed results as presented here might be useful for controlling a qutrit gated by electromagnetic fields or any ThLS described by the model (\ref{equ1}). 

For ThLSs as discussed here, we have also observed  coherent destruction of tunneling (CDT) when the longitudinal drive is tuned such that $A/\omega$ corresponds to one of the zeroes of the Bessel function. We have equally observed a sequence of consecutive $SU(3)$ LZSM oscillations when the amplitude of the longitudinal driving field widely exceeds the uniaxis anisotropy ($A\gg D$) and the frequency of the transverse drive is weak enough. We have reported a new type of  LZSM interference patterns that are observed when the transverse drive is tuned in the high-frequency regime and its frequency $\omega_{f}$ matches the easy axis anisotropy $D$. It is demonstrated that this can  be observed in crystal lattices with large $D$. 

Meanwhile, we have to stress that despite the success of our treatments in the weak and strong driving limits, some experiments might need to be performed in an intermediate regime i.e.  the regimes of moderated values of $A/\omega$ and $\mathcal{A}_{f}/\omega$ (not too large, not too small). For problems of control for instance, ThLSs might operate in the regime of moderate values. In such cases, we are afraid that our results might not be so relevant. This opens yet another perspective of using the model (\ref{equ1}) to address/explore this issue quite relevant for technological purposes. In this regard, one may think of deriving associated differential equations. This task unfortunately results into a class of third order Hill's equations in which solutions cannot be written in closed form. Therefore, the alternative method of transfer matrices could be of interest, but this is yet to be proven. We have to mention in conclusion that the effects of dissipative environment have not been considered. Another key and interesting question in this regard is about implications if the fields are quantized. 

\section*{Acknowledgments}

We are deeply indebted to Prof. M. N. Kiselev for his invaluable comments during the realization of the project in Ref.\onlinecite{Ken2013} that have deeply inspired the perspective of this work. We equally acknowledge valuable discussions with Prof. K. K. Kikoin on Gell-Mann matrices. We also appreciate discussions with Prof. Sigmund Kohler. 
M. B. Kenmoe thanks ICTP for the hospitality where this project started and the African Institute for Mathematical Science (AIMS) of Ghana for the warm hospitality during his stay and where this project was entirely written. We are deeply grateful  to F. Koyka, M. Nelson and J. Getz for careful reading of the manuscript and linguistic suggestions.

\appendix
\section{Eigen -values and -functions}\label{App1}
In this appendix, we present some supplemental ingredients of our adiabatic analysis. We consider arbitrary three-levels for which Rabi interactions between levels with extremal spin projections are set to zero and the intermediate state is off-resonance.  
\begin{eqnarray} \label{A0} 
\mathcal{H}(t)=\left[\begin{array}{ccc} {\omega _{+} (t)} & {\Delta _{12} (t)} & {0} \\ {\Delta _{21} (t)} & {0} & {\Delta _{23} (t)} \\ {0} & {\Delta _{32} (t)} & {\omega _{-} (t)} \end{array}\right]. 
\end{eqnarray}
$\omega _{\pm} (t)$ are detuning i.e. the difference between the Bohr transition frequency of the system and that of the external field, $\Delta _{ij} (t)$  couples the state $|i\rangle$ and $|j\rangle$. This model has been discussed in the main text in the adiabatic limit. We have barely shown that solutions strongly depend on 
eigen-values and eigen-functions. First quantities are given in Eq.(\ref{a2}) of the main text,   
where the functions $f_{\kappa j}(t)$ are given by:
\begin{eqnarray}\label{A2a}
f_{11}=\Delta_{12}\Delta_{23},\quad f_{12}=(E_{1}-\omega_{+})\Delta_{23}, 
\end{eqnarray}
\begin{eqnarray}\label{A2b} 
\hspace{-1cm} f_{13}=E_{1}(E_{1}-\omega_{+})-\Delta_{12}\Delta_{21},\quad  
f_{21}=(E_{2}-\omega_{-})\Delta_{12}, 
\end{eqnarray}
\begin{eqnarray}\label{A2c}
f_{22}=(E_{2}-\omega_{+})(E_{2}-\omega_{-}),  \quad f_{23}=(E_{2}-\omega_{+})\Delta_{32}, \quad
\end{eqnarray}
\begin{eqnarray}\label{A2}
f_{31}=\Delta_{21}\Delta_{32},\quad f_{32}=(E_{3}-\omega_{+})\Delta_{21},\slabel{A2d}
\end{eqnarray}
\begin{eqnarray}\label{A2}
f_{33}=E_{3}(E_{3}-\omega_{+})-\Delta_{23}\Delta_{32}, 
\end{eqnarray}
Here, $p(\tau)$ and $q(\tau)$ are respectively given by
\begin{eqnarray}\label{A3}
p=\frac{(\omega_{+}+\omega_{-})^{2}}{3}-\omega_{+}\omega_{-}+[\Delta_{12}\Delta_{21}+\Delta_{23}\Delta_{32}], 
\end{eqnarray}
and
\begin{eqnarray}\label{A4}
\nonumber q=\frac{2(\omega_{+}+\omega_{-})^{3}}{27}-\Delta_{12}\Delta_{21}\omega_{-}-\Delta_{23}\Delta_{32}\omega_{+}\\-
\frac{\Big(\omega_{+}\omega_{-}-[\Delta_{12}\Delta_{21}+\Delta_{23}\Delta_{32}]\Big)\Big(\omega_{+}+\omega_{-}\Big)}{3}. 
\end{eqnarray}


\begin{thebibliography}{99}
\bibitem{comment}
Such a problem leads to a class of third order linear differential Hill equation that cannot be solved in closed-form.
\bibitem{comment3}
Remark $\mathcal{H}_{Q}(t)$ and $\mathcal{H}_{drive}(t)$ have independent and different periods, $\mathcal{H}_{Q}(t+2\pi/\omega)=\mathcal{H}_{Q}(t)$ and $\mathcal{H}_{drive}(t+2\pi/\omega_{f})=\mathcal{H}_{drive}(t)$. Thus, at time $2\pi/\omega$ the phase difference between the two signals is $2\pi\omega_{f}/\omega$ while at time $2\pi/\omega_{f}$ the phase difference is $2\pi\omega/\omega_{f}$. If in the first case the fields are set such that $\omega\gg\omega_{f}$ and $\omega\ll\omega_{f}$ in the second case, there is no significant phase difference between both signals that act almost simultaneously.
\bibitem{comment5}
Obtaining Eq.(\ref{equ10}), we haven't assumed any fast oscillating exponential  in order to neglect it. Eq.(\ref{equ9}) is exact.
\bibitem{Comment2}
We have also found that $\bar{\omega}_{n}^{\pm}=\bar{\omega}_{n}$, $F^{\pm}_{\mu}(\tau)=F_{\mu}(\tau)$ with $\mu=cc, ss$,   our  analytical results acquire a simplified form and cascaded LZSM transitions are more visible
\end{thebibliography}
\end{document}